# Tracer Applications of Noble Gas Radionuclides in the Geosciences

(August 20, 2013)

Z.-T. Lu[a,b], P. Schlosser[c,d], W.M. Smethie Jr.[c], N.C. Sturchio[e], T.P. Fischer[f], B.M. Kennedy[g], R. Purtschert[h], J.P. Severinghaus[i], D.K. Solomon[j], T. Tanhua[k], R. Yokochi[e,l]

[a] Physics Division, Argonne National Laboratory, Argonne, Illinois, USA
[b] Department of Physics and Enrico Fermi Institute, University of Chicago, Chicago, USA
[c] Lamont-Doherty Earth Observatory, Columbia University, Palisades, New York, USA
[d] Department of Earth and Environmental Sciences and Department of Earth and Environmental Engineering, Columbia University, New York, USA
[e] Department of Earth and Environmental Sciences, University of Illinois at Chicago, Chicago, IL, USA
[f] Department of Earth and Planetary Sciences, University of New Mexico, Albuquerque, USA
[g] Center for Isotope Geochemistry, Lawrence Berkeley National Laboratory, Berkeley, USA
[h] Climate and Environmental Physics, Physics Institute, University of Bern, Bern, Switzerland
[i] Scripps Institution of Oceanography, University of California, San Diego, USA
[j] Department of Geology and Geophysics, University of Utah, Salt Lake City, USA
[k] GEOMAR Helmholtz Center for Ocean Research Kiel, Marine Biogeochemistry, Kiel, Germany
[l] Department of Geophysical Sciences, University of Chicago, Chicago, USA

## Abstract

Noble gas radionuclides, including $^{81}$Kr ($t_{1/2}$ = 229,000 yr), $^{85}$Kr ($t_{1/2}$ = 10.8 yr), and $^{39}$Ar ($t_{1/2}$ = 269 yr), possess nearly ideal chemical and physical properties for studies of earth and environmental processes. Recent advances in Atom Trap Trace Analysis (ATTA), a laser-based atom counting method, have enabled routine measurements of the radiokrypton isotopes, as well as the demonstration of the ability to measure $^{39}$Ar in environmental samples. Here we provide an overview of the ATTA technique, and a survey of recent progress made in several laboratories worldwide. We review the application of noble gas radionuclides in the geosciences and discuss how ATTA can help advance these fields, specifically: determination of groundwater residence times using $^{81}$Kr, $^{85}$Kr, and $^{39}$Ar; dating old glacial ice using $^{81}$Kr; and an $^{39}$Ar survey of the main water masses of the oceans, to study circulation pathways and estimate mean residence times. Other scientific questions involving deeper circulation of fluids in the Earth's crust and mantle are also within the scope of future applications. We conclude that the geoscience community would greatly benefit from an ATTA facility dedicated to this field, with instrumentation for routine measurements, as well as for research on further development of ATTA methods.

# Contents

# 1. Introduction

Due to their inert behavior in natural systems, noble gas nuclides have been used for many decades in the earth sciences (Porcelli *et al.*, 2002). They provide fundamental information on mass budgets and exchange, as well as mean residence times in the hydrosphere (oceans, lakes, and groundwater), the atmosphere, the geosphere, and the cryosphere. While the stable nuclides of noble gases can be measured routinely with high precision using mass spectrometry, the radioactive nuclides of noble gases in natural samples are rare and very difficult to measure with sufficient precision. The radionuclides, including $^{81}$Kr (half-life, $t_{1/2}$ = 229,000 yr), $^{85}$Kr ($t_{1/2}$ = 10.8 yr), and $^{39}$Ar ($t_{1/2}$ = 269 yr), possess nearly ideal chemical and physical properties for studies of earth and environmental processes. In practice, however, few studies have been able to utilize these tracers because of the large sample sizes and the complex analytical systems required for measuring their extremely low isotopic abundances ($10^{-11}$ – $10^{-16}$). Low-level radioactive decay counting is applied to the analysis of $^{39}$Ar and $^{85}$Kr, but is not possible for $^{81}$Kr due to its low activity. Various atom-counting methods have been pursued over the past four decades, including efforts focused on Accelerator Mass Spectrometry (AMS) (Collon *et al.*, 2004).

The advent of ATTA (Atom Trap Trace Analysis) (Jiang *et al.*, 2012; Yang *et al.*, 2013) in principle enables routine measurements of rare noble gas isotopes and thus has reignited the discussion among Earth scientists about how to best apply these valuable tracers to compelling scientific problems. In the first Workshop on Tracer Applications of Noble-Gas Radionuclides (TANGR2012), held at Argonne National Laboratory in June, 2012, the attendees reviewed past applications of these tracers in the field of Earth science and projected future needs. The scientific findings are summarized in this article. Additional materials are available at the workshop website, http://www.phy.anl.gov/events/tangr2012/.

After a brief overview of the relevant noble gas radionuclides, we describe the ATTA method and specific experimental instruments. We then discuss suitable methods for noble gas sample collection and purification. These method- and technique-oriented chapters are followed by a discussion of the applications of $^{85}$Kr, $^{39}$Ar and $^{81}$Kr in the hydrosphere (oceans and groundwater), cryosphere and geosphere. The Conclusions section projects the trajectories for the field of applying noble gas radionuclides to earth and environmental sciences, and recommends concrete pathways towards more widespread use of these isotopes in the geosciences.

## 2. Noble Gas Radionuclides

There are three long-lived ($t_{1/2}$ > 1 yr) noble-gas radionuclides with tracer applications in the environment: $^{81}$Kr ($t_{1/2}$ = 229,000 yr), $^{85}$Kr (10.8 yr), and $^{39}$Ar (269 yr). Being chemically inert, these three nuclides predominantly reside in the atmosphere. They follow relatively simple mixing and transport processes in the environment, and they can be extracted from large quantities of water or ice for analysis. These geophysical and geochemical properties are favorable for the purpose of radio-isotope dating (Collon *et al.*, 2004). The half-lives of the three tracer nuclides have different orders of magnitude, allowing them to cover a wide range of ages (Fig. 1).

$^{81}$Kr is a cosmogenic nuclide with an atmospheric $^{81}$Kr/Kr ratio of $(5.2 \pm 0.4) \times 10^{-13}$ (Collon *et al.*, 1997). It has a long residence time and a spatially homogeneous distribution in the atmosphere, making it a desirable tracer for its dating range (Loosli and Oeschger, 1969).

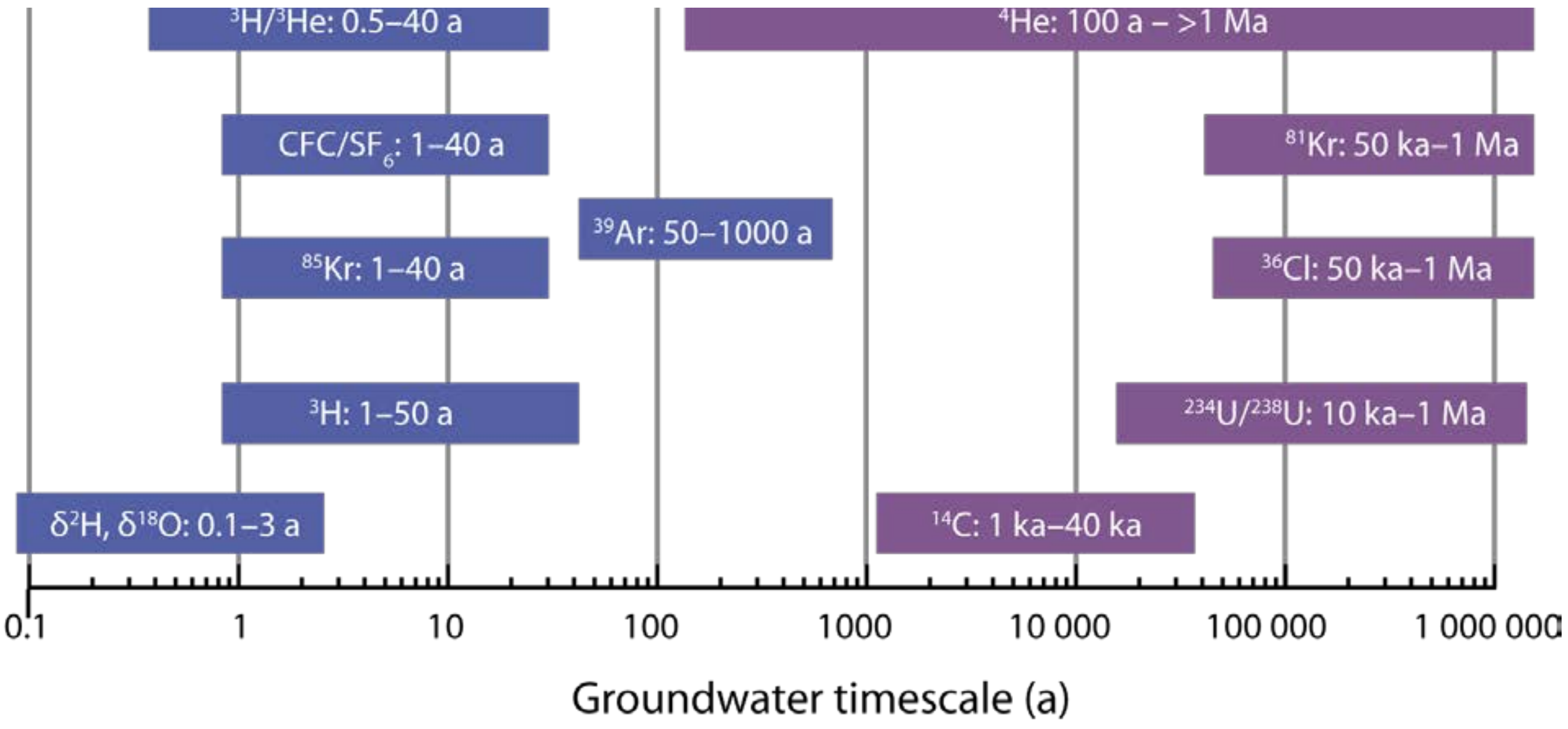

**FIG. 1.** Age ranges of radioisotope dating using $^{85}Kr$, $^{39}Ar$, $^{81}Kr$ and other established environmental tracers. Figure reproduced from Aggarwal, 2013.

Compared to atmospheric production of $^{81}Kr$, contributions from both spontaneous and neutron-induced fission are negligible because $^{81}Kr$ is shielded by $^{81}Br$ from the fission yields on the neutron-rich side (Lehmann *et al*., 1993). At present, ATTA is the only method capable of measuring $^{81}Kr/Kr$ in environmental samples. The required sample size for a typical analysis is 100 – 200 L of water or 40 – 80 kg of ice.

$^{85}Kr$ is amply produced during fission in nuclear reactors, and is released into the atmosphere due to reprocessing of spent nuclear fuel rods (Turkevitch *et al*., 1997; Ahlswede *et al*., 2013). The atmospheric $^{85}Kr/Kr$ ratio is approximately $2\times10^{-11}$. Due to its relatively short half-life, the spatial distribution of $^{85}Kr$ in the atmosphere is not as uniform as that of $^{39}Ar$ or $^{81}Kr$. For example, $^{85}Kr/Kr$ in the northern hemisphere, where most of the nuclear fuel reprocessing plants reside, can be ~ 20% higher than that in the southern hemisphere (Weiss *et al*., 1989). $^{85}Kr$ is useful for dating young (1 – 40 yr) groundwater (Smethie Jr. *et al*., 1992; Althaus *et al*., 2009; Momoshima *et al*., 2011). At present, both Low-Level Decay Counting (LLC) and ATTA can measure $^{85}Kr$ in environmental samples. The results from the two methods agree in comparison studies (Jiang *et al*., 2012; Yang *et al*., 2013).

$^{39}Ar$ conveniently fills an apparent age gap (Fig. 1) between $^{85}Kr$ and $^3H/^3He$ on the shorter and $^{14}C$ on the longer time scale. This makes $^{39}Ar$ a much desired isotope for dating environmental samples on the time scale of a few hundred years (Loosli and Oeschger, 1968; Lehmann and Purtschert, 1997). Atmospheric $^{39}Ar$ is of cosmogenic origin with a $^{39}Ar/Ar$ ratio of $8\times10^{-16}$ (Loosli, 1983). There can be substantial subsurface production in granite rocks

through the $^{39}K(n, p)^{39}Ar$ reaction and muon capture on $^{39}K$ (Lehmann *et al.*, 1993; Mei *et al.*, 2010). This effect should be evaluated in $^{39}Ar$ dating of groundwater, particularly in geothermal systems or fractured crystalline rocks, for example, by simultaneous measurements of $^{37}Ar$ ($t_{1/2}$ = 35 d) produced by neutron-activation of $^{40}Ca$. In ocean water, *in situ* production of $^{39}Ar$ is not significant. However, atmospheric $^{39}Ar$ dissolved in ocean water is a useful chronometer for tracing ocean circulation and ventilation. At present, LLC is used for $^{39}Ar$ (and $^{37}Ar$) analysis, but it requires 1-2 tons of water per sample. The ATTA instruments have been used to measure $^{39}Ar/Ar$ in environmental samples on much smaller amounts of Ar than those required for LLC (Jiang *et al*., 2011).

# 3. Atom Trap Trace Analysis

Atom Trap Trace Analysis (ATTA) is a laser-based atom counting method (Chen *et al*., 1999). A magneto-optical trap is used to capture atoms of the desired isotope in the center of a vacuum chamber using laser beams (Fig. 2). A photo-sensor detects the laser induced fluorescence emitted by the trapped atoms. Trapping force and fluorescence detection require the atom to repeatedly scatter photons at a high rate (~$10^7$ $s^{-1}$), a process that is the key to the superior selectivity of ATTA because it only occurs when the laser frequency precisely matches the resonance frequency of a particular atomic transition. Even the small changes in the atomic transition frequency between isotopes of the same element, the so called isotope shifts caused by changes in nuclear size and mass, are sufficient to perfectly distinguish between the isotopes. ATTA is unique among trace analysis techniques in that it is free of interferences from any other isotopes, isobars, atomic or molecular species.

## 3.1. Argonne National Laboratory

Following the first demonstration of ATTA (Chen *et al.*, 1999), both the reliability and counting efficiency of the instruments have been steadily improved. With the completion of the ATTA-3 instrument at Argonne (Jiang *et al.*, 2012), for the first time $^{81}Kr$-dating became available to the geoscience community at large. In the period since November 2011, the Argonne group has measured both $^{81}Kr/Kr$ and $^{85}Kr/Kr$ ratios in ~ 50 samples that had been extracted by collaborators from groundwater wells in the Great Artesian Basin, Australia (Love *et al*., 2012), Guarani Aquifer, Brazil (Aggarwal *et al*., 2012), and Locust Grove, Maryland; from brine wells of the Waste Isolation Pilot Plant, New Mexico (Sturchio *et al*., 2013); from geothermal steam vents in the Yellowstone National Park (Yokochi *et al*. 2013); and from near-surface ice at Taylor Glacier, Antarctica (Buizert *et al*., 2013). Sample purification was conducted by either the University of Illinois at Chicago or University of Bern (see Section 4).

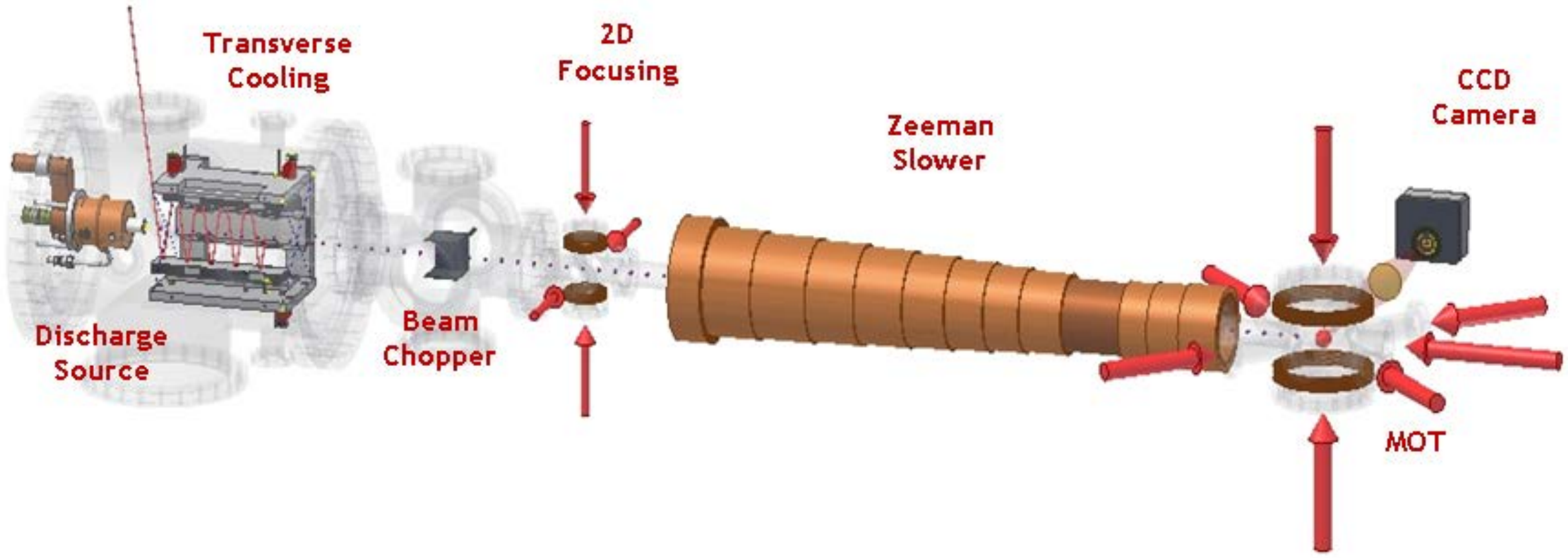

**FIG. 2.** Schematic of the ATTA-3 apparatus developed at Argonne National Laboratory. The total length of the atomic beamline is approximately 2 m. Lasers and optics are located on an adjacent laser table of a similar length. Figure reproduced from Jiang *et al*., 2012.

The required sample size for applications in $^{81}$Kr dating depends on both the sample age and the desired uncertainty in age determination (Fig. 3). $^{81}$Kr-dating with ATTA-3 covers an effective age range from 150 kyr to 1.5 Myr, or 0.6 – 6 times the half-life of the isotope. On the side younger than 150 kyr, the change of $^{81}$Kr/Kr is too small to provide precise age resolution. On the side older than 1.5 Myr, the $^{81}$Kr/Kr ratio itself is comparable to the error introduced by the correction for a memory effect related to ion implantation (discussed below). Within the effective age range, a typical sample size is 5 – 10 micro-L STP of krypton, which can be extracted from approximately 100 – 200 L of water or 40 – 80 kg of ice. (Kr has a higher concentration in ice than in water due to the gas bubbles trapped in ice.) For an $^{85}$Kr/Kr analysis, the required sample size is generally smaller by an order of magnitude because of the isotope's higher initial abundance in the atmosphere. It should be noted that these are not absolute requirements. Instead, they should be viewed as a guideline. If needed, extraordinary steps, for example prolonged xenon flushing in order to reduce the memory effect, can be taken to further reduce the required sample size and meet the special demands of a particular application. The chemical purity of the krypton sample is not critical since the ATTA method is immune to contamination from any other species.

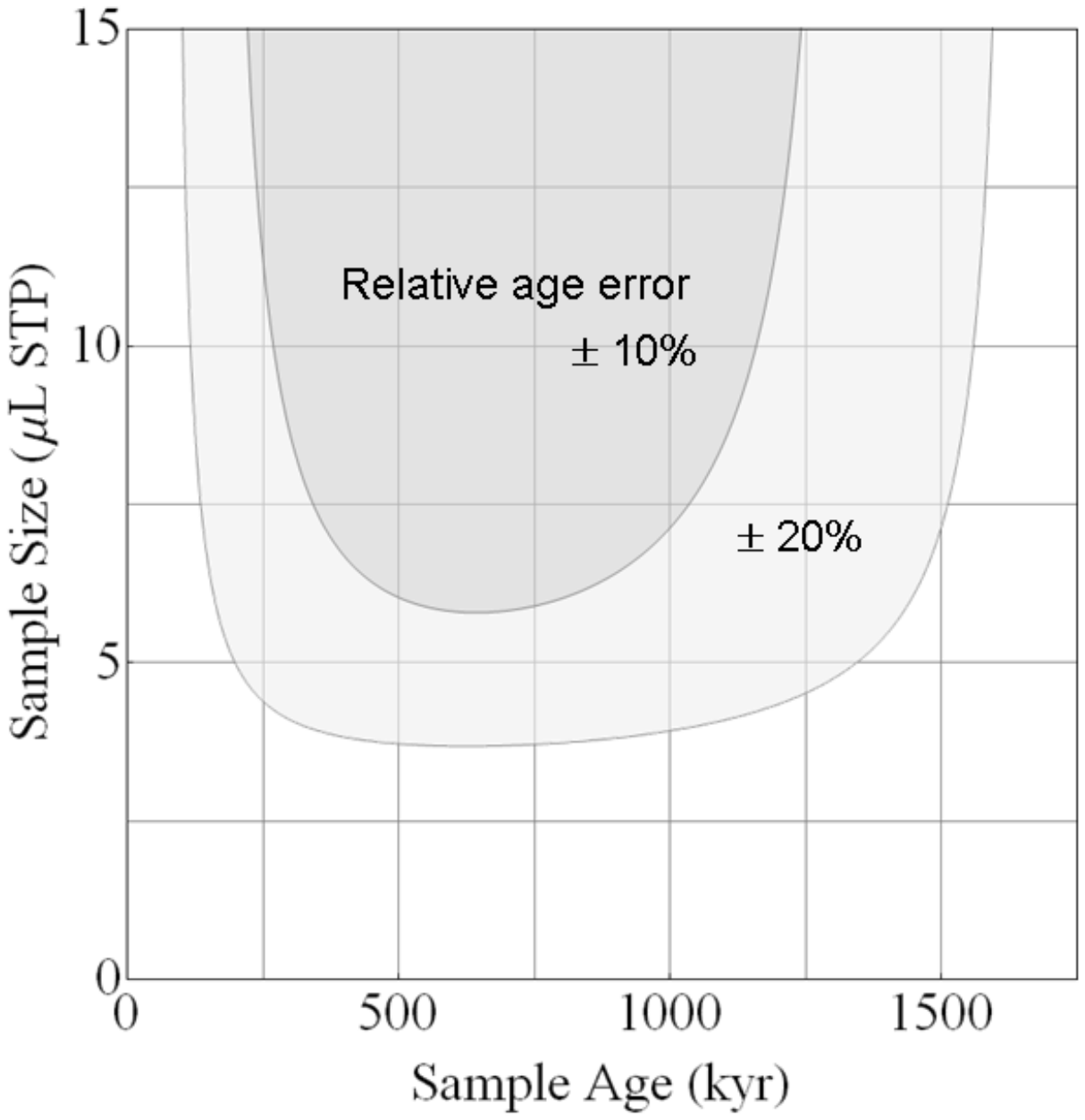

**FIG. 3.** Sample size vs. sample age and desired accuracy for $^{81}$Kr-dating. The two curves are for a relative age error of ±10% and ±20%, respectively. Figure reproduced from Jiang *et al*., 2012.

Cross-sample contamination remains the primary limitation to the sample size requirement and sample processing time of ATTA-3. The discharge used to excite krypton atoms causes ionization and the implantation of ions into the surrounding walls. Those embedded atoms of the current and previous samples can later be gradually released back into the vacuum system, causing an instrumental memory effect. This effect is mitigated by flushing the system for 36 hours with a xenon gas discharge following each measurement. Discharges of noble gases and hydrogen were investigated for this task, and the one based on xenon was found to be the most effective, perhaps due to its large atomic mass. This solution significantly limits the sample processing speed as each measurement-flushing cycle takes two days of time even though the measurement itself only takes 2-4 hours. Taking into account the need to calibrate the system regularly with a standard sample, the Argonne group can presently analyze $^{81}$Kr/Kr and $^{85}$Kr/Kr in 120 samples annually. Additional atom traps sharing a common laser system can be introduced to increase the sample throughput at the rate of 120 samples per year per trap.

With future R&D, the flushing time of 36 hours could be shortened. Moreover, the discharge source of metastable atoms may be replaced with a photon excitation scheme (Ding *et al*., 2007), thus avoiding the undesirable effects due to ionization in the discharge. If successful, the photon excitation scheme will lead to even smaller required sample size and a higher sample processing speed. With photon excitation, the trapping and counting efficiency will be further improved, and the 36-hour flushing will no longer be necessary.

A demonstration experiment was conducted at Argonne in 2010 to show that ATTA-3 can analyze $^{39}$Ar/Ar ratios in both an atmospheric sample at the level of $8\times10^{-16}$ and an old water sample at the level of $1\times10^{-16}$ (Jiang *et al*., 2011). Indeed no interference from other atomic or molecular species was observed at the $1\times10^{-16}$ level. Counting statistics was the only limitation. The counting rate for the atmospheric sample was only 6 counts per day. It should be noted that the size of a typical argon gas sample extracted for ocean ventilation studies, about 10 mL-STP from 30 liters of water, appears adequate for an ATTA analysis.

## 3.2. University of Science and Technology of China

The group at the University of Science and Technology of China (USTC) has collaborated with the Argonne group to develop both the ATTA-3 instrument at Argonne and an ATTA instrument at USTC (Cheng *et al*., 2010; Yang *et al*., 2013). There is a major difference between the two instruments in that the loading rate of the control isotope ($^{83}$Kr) is measured by two different methods. Compared to single-atom counting of $^{81}$Kr or $^{85}$Kr, an accurate determination of the trap capture rate of the abundant isotope $^{83}$Kr is surprisingly difficult, yet it is required in order to measure the isotopic abundances of $^{81}$Kr/Kr and $^{85}$Kr/Kr. Because the loading rate of $^{83}$Kr is higher than those of the rare isotopes by more than 10 orders of magnitude, interaction among the large number (~$10^9$) of $^{83}$Kr atoms in the trap causes significant nonlinear effects in the measurement of its high loading rate. The ATTA-Argonne instrument employs a continuous quench and fluorescence monitoring scheme (Jiang et al., 2012), while the ATTA-USTC instrument employs a pulsed quench-and-capture scheme (Cheng *et al*., 2013).

With both ATTA systems operating, an inter-comparison study was carried out to verify the reliability and reproducibility of the results. The $^{85}$Kr/Kr ratios of a group of 12 samples, in the range of $10^{-13}$ to $10^{-10}$, were measured independently in three laboratories: a low-level counting (LLC) laboratory in Bern, Switzerland, and the two ATTA laboratories, at USTC and Argonne. The results are in agreement at the precision level of 5% (Fig. 4).

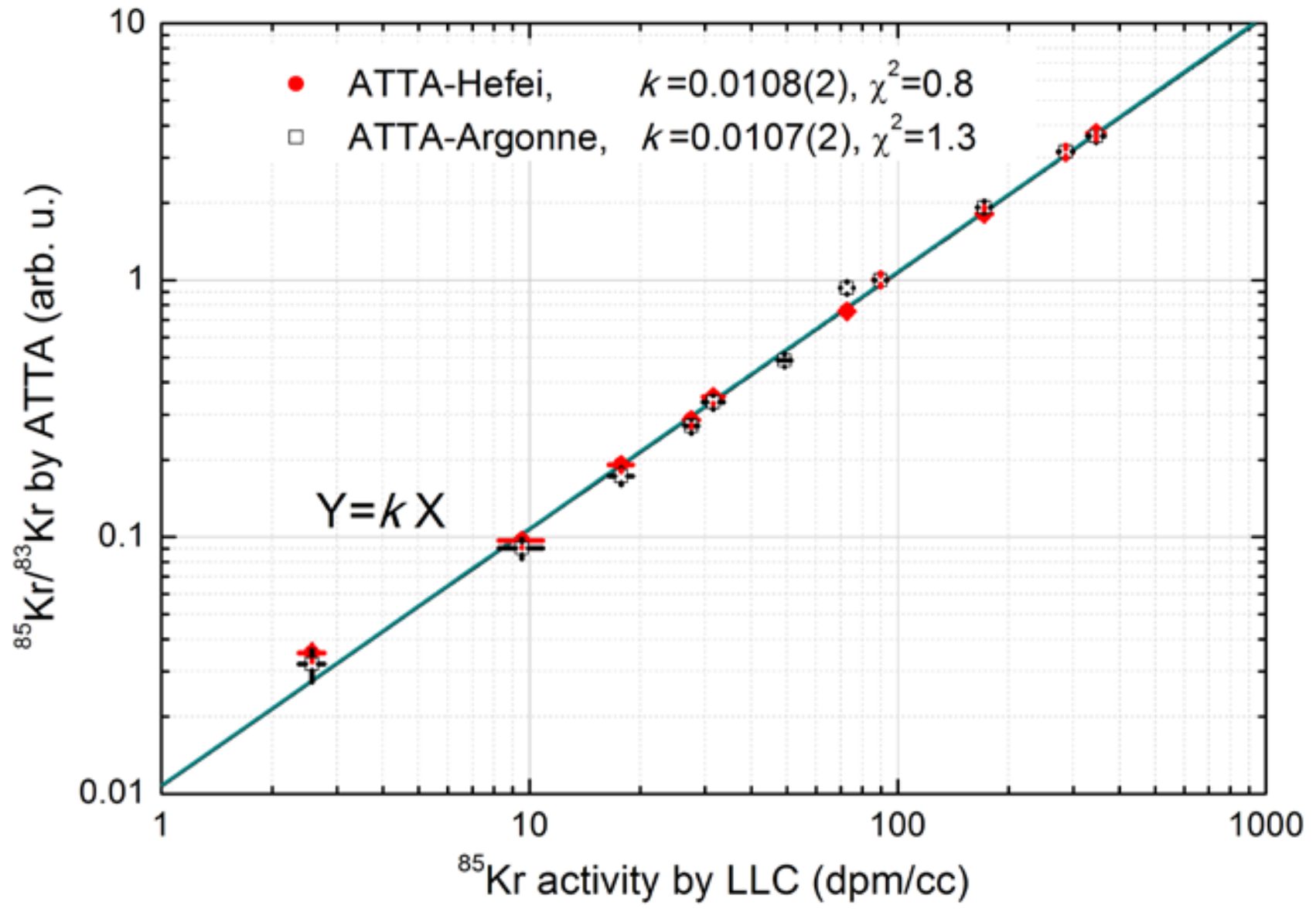

**FIG. 4.** Comparison between the $^{85}$Kr/$^{83}$Kr ratios determined by ATTA-USTC (solid red circles), ATTA-Argonne (open black squares) and the $^{85}$Kr activity by LLC (in the units of decays per minute in 1 cm$^3$-STP of Kr gas). The data are fitted with proportional functions, and the proportional coefficients (*k* values) are given for both data sets. Figure reproduced from Yang *et al*., 2013.

## 3.3. University of Heidelberg

The Heidelberg group is focused on developing an ATTA system for $^{39}Ar/Ar$ analysis (Welte *et al.*, 2010). The group demonstrated an $^{39}Ar$ atom count-rate of 4.1 ± 0.3 $hr^{-1}$ for atmospheric samples, which represents an improvement by a factor of 18 over what the Argonne group had demonstrated earlier. With only one day of data acquisition, the Heidelberg group dated a groundwater sample to be 360 ± 68 years. The instrument has the potential to measure $^{39}Ar/Ar$ ratios in samples of 1 cc STP of argon, which can be extracted from 100 cc of air, 2.5 liters of water, or 1 kg of ice (Ritterbusch *et al.*, 2013).

Groundwater samples from the Upper Rhine Graben aquifers were collected and analyzed by established methods for a large range of tracers, including tritium, stable isotopes, noble gases, and $^{14}C$. For $^{39}Ar$ analysis, several tons of water was degassed in the field using a membrane contactor. In the laboratory, a gas-chromatographic system at cryogenic temperatures was used to separate pure argon from the extracted gas. ATTA was then used to isolate and count $^{39}Ar$ atoms from these samples. In parallel, samples for $^{39}Ar$ analysis by LLC at the University of Bern were taken to enable comparison of the two analytical techniques.

The resulting $^{39}Ar$ groundwater ages in the range of several hundred years are in accordance with the indications obtained from the classical age tracers (Reichel *et al.*, 2013). They provide quantitative information on the groundwater travel time for an important, strongly exploited part of the investigated aquifer system, which could not be obtained from the other tracers. These results significantly improve the knowledge of the time scale of groundwater renewal in the aquifer layers of intermediate depth. Furthermore, the combination of the $^{39}Ar$ age scale with noble gas recharge temperatures and stable isotope data has the potential to provide a reconstruction of the regional paleoclimate on the time-scale of the past millennium.

## 3.4. Columbia and Hamburg Universities

A group in the Physics Department of Columbia University is developing an ATTA system for the purpose of measuring $^{84}Kr/Xe$ ratios down to the $10^{-12}$ level, not in environmental samples, but in highly purified xenon gas samples (Aprile *et al.*, 2013). $^{84}Kr$ is stable and has a $^{84}Kr/Kr$ ratio of 57% in terrestrial samples. For a large-volume (> 100 L) liquid xenon detector of dark matter, the radioactive decays of the $^{85}Kr$ impurities remain a limitation to the detector sensitivity. While cryogenic distillation is an established technology for removing Kr (including both $^{85}Kr$ and $^{84}Kr$) from Xe, there needs to be a method to determine the level of $^{85}Kr$ contamination. In atmospheric samples, $^{85}Kr/Kr \sim 1\times10^{-11}$, consequently the desired contamination level of $^{85}Kr/Xe$ ($\sim 10^{-23}$) corresponds to a $^{84}Kr/Xe$ ratio of $\sim 10^{-12}$. An ATTA system has been constructed and tested with Ar, not Kr, in order to avoid Kr contamination. The system efficiency determined with Ar leads to an expected $^{84}$Kr/Xe sensitivity at the targeted level of $10^{-12}$.

A group at Hamburg University is developing an ATTA system to monitor atmospheric $^{85}$Kr released by nuclear fuel reprocessing plants. If employed by the International Atomic Energy Agency (IAEA), such a system could detect non-compliance to the Non-Proliferation-Treaty (Winger *et al.*, 2005). While the trap is under construction, the group has developed a vacuum ultraviolet (VUV) lamp that can excite krypton into the metastable state, and operated the lamp for over 500 hours (Daerr *et al.*, 2011).

# 4. Sampling

## 4.1. Sample Collection

Specialized methods for sampling and noble gas separation have been developed to satisfy the requirements of the relatively large volumes of argon and krypton needed for noble gas radionuclide measurements of gas, water, and ice. Earlier studies of $^{81}$Kr have used gases obtained from 17 tons (Lehmann *et al*., 2003) and 3 tons (Sturchio *et al*., 2004) of water, respectively. Current methods for analysis of $^{39}$Ar and $^{81}$Kr still require gas extraction in the field. Such gas extraction systems must have the following characteristics: (i) be leak tight to avoid sample contamination by modern air, (ii) have a high extraction yield (> 80%) to minimize the necessary volume of water and extraction time, and (iii) be sufficiently robust for field work, i.e. simple, light and portable (Purtschert *et al*., 2013a). Efficiency optimization also depends on whether the extraction is carried out on land or on a research ship (Smethie Jr. and Mathieu, 1986).

Gases can be extracted by exposing the air-saturated water to a gas phase having lower partial pressures of argon and krypton, which can be achieved either by applying a partial vacuum or by putting the water in contact with a pure and inert gas such as helium or $N_2$ in which argon and krypton partial pressures are nominally zero. The efficiency of gas extraction from groundwater, therefore, depends on (i) partial pressures of argon and krypton in the gas phase, (ii) solubility of argon and krypton in the water, and (iii) rate of gas transfer. The pressure in the gas phase depends on the capability of the pump, and on the gas flux, which is proportional to the water flux and the gas content of the water. The limiting factor of the gas transfer rate from water to the gas phase is the rate of gas diffusion in water. Increasing the specific gas-water interfacial surface area, either by spraying water or by using a membrane contactor characterized by a large surface area, can effectively enhance the efficiency of the apparatus.

During these gas extraction processes, the elemental and isotopic ratios of noble gases are expected to fractionate depending on their respective solubilities. A theoretical limit on the maximum possible isotopic fractionation during gaseous diffusion is given by assuming an ideal Rayleigh fractionation process (Broecker and Oversby, 1971), after using Graham's Law ([$v_1/v_2 = (m_2/m_1)^{0.5}$], where $v$ and $m$ are the velocity and mass, and subscripts *1* and *2* refer to light and heavy isotopes, respectively) to estimate the isotopic fractionation factor. For an extraction efficiency of >60%, the maximum isotopic fractionation of the $^{81}Kr/^{84}Kr$ ratio is <1%, and is much smaller than the uncertainty of the current measurements.

Two categories of systems have been applied in the field for large-volume gas extraction (Fig. 5): (i) vacuum extraction chamber and (ii) membrane contactors. In both systems, water is transferred from the well or the sampler into the system and discharged after gas extraction. Large particles in the water are filtered out, and water flux and inlet pressure are monitored using a flow meter and a pressure gauge. Check valves are used to prevent water or gas from flowing in undesired directions.

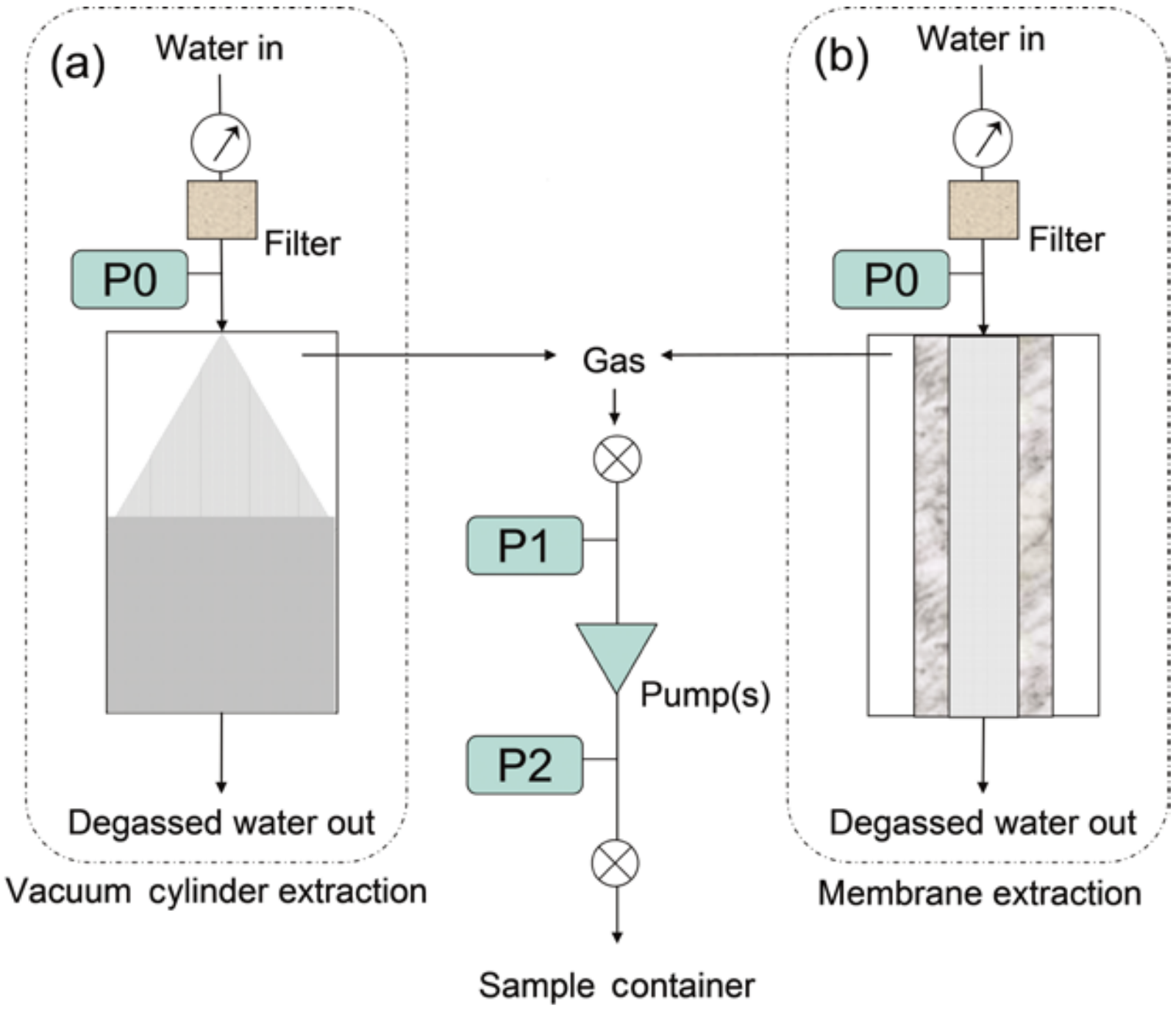

**FIG. 5.** Schematic diagram of water degassing systems: a) vacuum cylinder extraction; b) membrane extraction. (P0: water inlet pressure; P1: gas extraction pressure; P2: gas pressure in the sample container). Figure reproduced from Purtschert *et al*., 2013a.

In a vacuum extraction chamber, water is sprayed through nozzles of optimal size (2-6 mm) into a transparent cylinder containing two phases, gas and water, separated by gravity. Water is continuously removed from the bottom of the cylinder at a desired rate regulated with magnetic floating sensors to maintain a range of water levels, while gas is transferred to a sample container through a compressor. This method has the advantage of being insensitive to temperature variation and to chemical contaminants in the sample water. For air-saturated water with a flux of 20 L min$^{-1}$, a typical extraction pressure of 20-50 mbar is reached. This leads to an extraction efficiency of 80-90% for water at 10°C (Kropf, 1996). The flow rate, and thus the sampling capacity of the system, is primarily limited by the rate of the drainage pump, at about 30 L min$^{-1}$. Vacuum extraction systems developed at the Physics Institute, University of Bern, have been used successfully in field campaigns for large volume $^{39}$Ar, $^{85}$Kr, and $^{81}$Kr sampling (Lehmann *et al*., 2003; Sturchio *et al*., 2004; Corcho *et al*., 2007). Similar systems have also been built at University of Heidelberg (Schlosser *et al*., 1989), University of Freiburg, and Hydroisotop GmBH for research in hydrology and oceanography (Smethie Jr. and Mathieu, 1986).

In a hydrophobic, microporous, hollow-fibre membrane contactor, there are two cylindrical volume sections partitioned by the membrane. Being hydrophobic, the membrane does not allow liquid water to pass through the pores, and diffusion of gas from the liquid water into the gas-filled pores is the mechanism by which gases are extracted from water (Wiesler, 1996). The gas phase in contact with water is maintained at a pressure of 20-150 mbar (P1, Fig. 5), and the gas released from the water is pumped into sample containers (P2). The commercially available membranes facilitated the construction of compact field degassing systems (Ho *et al*., 2002; Probst *et al*., 2007; Ohta *et al*., 2009). The membrane contactor currently used at the University of Illinois at Chicago (Liqui-Cel® Extra-Flow 4 x 28) handles water flow rates up to 30 L min$^{-1}$ and water temperatures to 70°C. For air-saturated water with a water flux of 20 L min$^{-1}$ at a typical vacuum pressure of 130 mbar, the extraction efficiency is 80-90% for argon and $O_2$, and >70% for krypton. No significant krypton isotopic fractionation has been observed through this extraction process. The operating pressure of the vacuum is limited by the continuous flux of gas through the membrane and the rate at which the compressor can transfer the gas into the sample container. The vacuum-side pressure is normally kept at 50-100 mbar, which in turn limits the extraction efficiency. The original prototype system, named EDGAR (Extraction of Dissolved Gases for Analysis of Radiokrypton), weighed about 180 kg (Probst *et al*., 2007). A recent modification (EDGAR-2) decreased the size and has been replicated and modified by groups in Vienna (IAEA), Brazil, and Heidelberg. Smaller membrane contactors are now available with lower water flow capacity (< 5 L min$^{-1}$), and may be applied for sampling smaller volumes (Ohta *et al*., 2009).

## 4.2. Sample Preparation

Relatively large quantities (5-10 µL) of pure krypton gas are required for measurements of $^{81}$Kr by ATTA, and even larger quantities (~ 250 mL) of argon are required for low-level counting of $^{39}$Ar. The gas phase extracted in the field, with a highly variable composition, has to be processed and purified in the laboratory to produce suitably pure argon and krypton for analyses. There are several methods to separate noble gases from gas mixtures. Some of the most common processes used in large-volume Kr-Ar purification system are distillation, adsorption, and absorption.

New gas purification techniques and materials (e.g., membranes, nano-tubes, and cation-exchanged zeolites) have recently become available. As the efficiency of ATTA measurements increases and required sample size decreases, separation and purification techniques will converge toward techniques that are presently in use for stable noble gas isotopic analyses (Beyerle *et al.*, 2000). The procedures described here or elsewhere in the literature (Smethie Jr. and Mathieu, 1986; Loosli and Purtschert, 2005; Yokochi *et al.*, 2008) are therefore to be regarded as a snapshot of the present capability. Two large-volume noble gas purification systems are presented below.

### 4.2.1. University of Bern: argon and krypton

The Physics Institute of the University of Bern has a long history in the separation and analysis of noble gas radionuclides by low-level counting (LLC) ( Loosli, 1983; Loosli *et al.*, 1986; Loosli *et al.*, 1999; Loosli and Purtschert, 2005). Recently, a more efficient argon purification system was developed (Riedmann, 2011). In this system (Fig. 6), argon is separated in a one-step desorption process from a lithium-cation exchanged faujasite zeolite. The gas is first dried in a cryogenic trap at $LN_2$ temperature, then transferred into the system and adsorbed at - 120°C – - 130°C. The number of connected columns is adjusted to fit the sample size up to a maximal volume of 110 litres of air-like gas when all nine columns are used. A complete gas transfer to the separation columns is achieved by means of a rotary pump, which evacuates the sample flask to < 10 mbar. The system is then purged with 6-7 litres of helium at 1400 mbar and a closed helium flux of ~8 L min$^{-1}$. During this procedure the columns are bypassed via valve 6 (V8 is closed). The separation process is started when the helium cycle is diverted though the GC columns (close V6 and open V8). Argon desorbs first within 10-20 minutes and is collected in an activated charcoal trap at $LN_2$ temperature (a.c. in Fig 6). The composition of the released gas is monitored by a quadrupole mass spectrometer (MS). The Ar fraction is purified in a getter module at an operating temperature of 400°C. The Ar separation yield is > 95% with a purity of better than 99.9%. The optimal separation temperature is a trade-off between the separation time, the Ar recovery efficiency, and the oxygen/argon separation performance. For example, gas from oxygen-free groundwater is separated at higher temperatures (-120°C) than air-like gas mixtures. The temperature is regulated in a vessel by cooling in a $LN_2$ atmosphere and by counter heating of the nine individual stainless-steel columns (total heating power 360 W).

After the Ar separation the columns are heated to 250 °C and regenerated. The desorbed gas is recycled and collected for further Kr purification (see below) for $^{85}$Kr-$^{81}$Kr analyses. Radon and Kr separate efficiently and the cross-sample memory effect has been tested to be less than 0.01%.

Compared to the processing time using the previous setup (2-3 days), argon from 100 litres of air can now be separated within 3-4 hours including the regeneration of the columns.

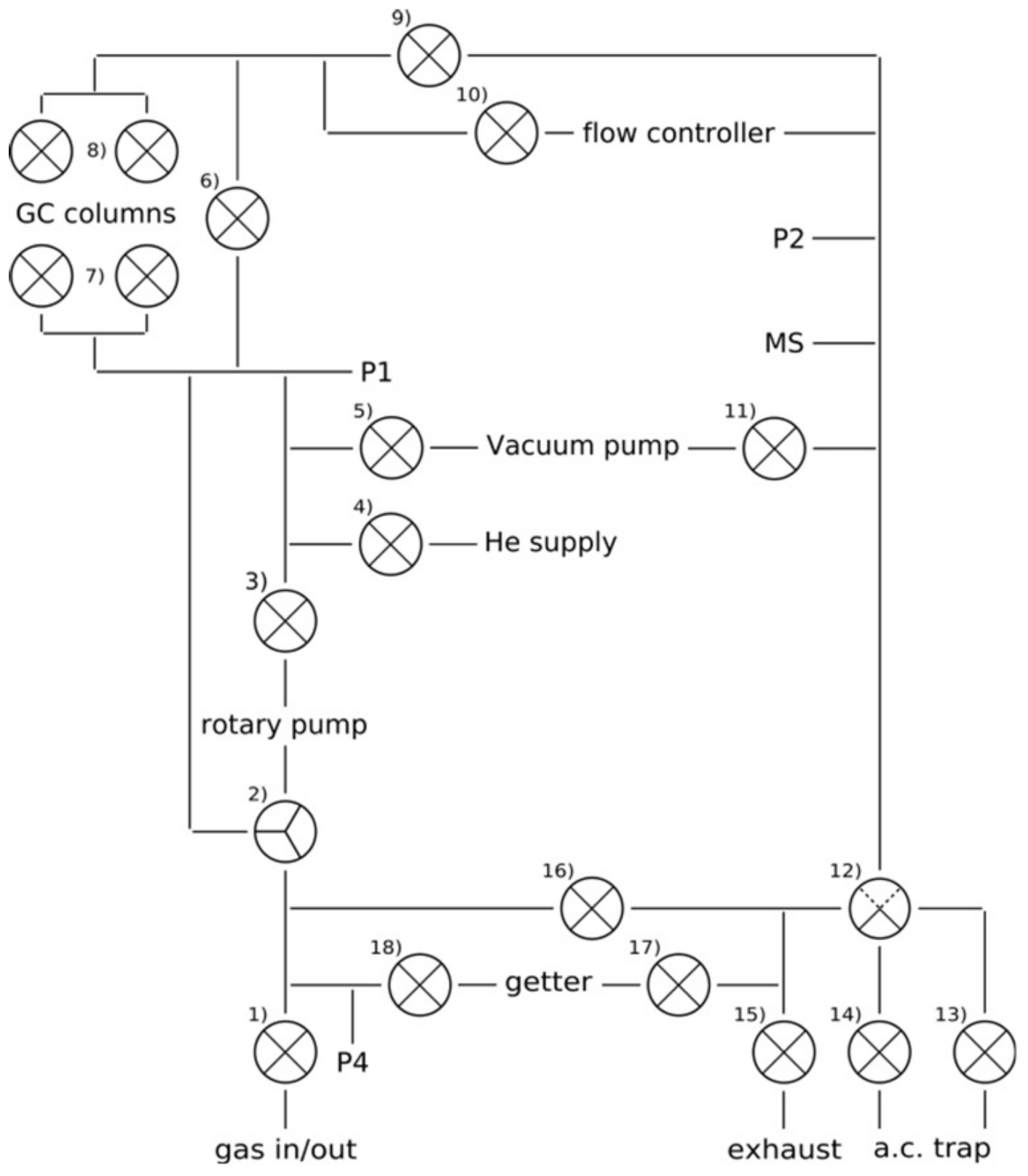

**FIG. 6.** Ar purification system at University of Bern. The crude gas sample is adsorbed on large GC columns and Ar is separated in a single-step desorption process by flushing with a He carrier at 153 K. A getter is used for the final purification. Figure reproduced from Riedmann, 2011.

Krypton from typically 300-500 liters of water, or from the aforementioned Ar-Kr separation system is further separated by three GC steps (Purtschert *et al*., 2013a). The sample gas is exposed to a molecular sieve 5-Angstrom bed to remove water vapor and $CO_2$, and subsequently is adsorbed on an AC (activated charcoal) cold finger at $LN_2$ temperature. The trap is then warmed to the room temperature, and the desorbed gases (mainly $N_2$ and Ar) are pumped away during the first five minutes. The trap is then flushed with helium carrier gas and the eluent is monitored with a TC detector. After the Ar, $O_2$ and $N_2$ peaks have passed, two-way valves are switched to divert the gas flow through the second trap filled with 5-Angstrom molecular sieve cooled at $LN_2$ temperature. From this trap, the desorption procedure is repeated to further purify and freeze Kr in the third cold trap filled with AC. In each step the gas volume is reduced by a factor of ~10 (Collon *et al*., 2000). The third cooling trap is then removed from the system and connected to a commercial GC system (Varian Star 3400) where Kr is further purified. If the Kr and $CH_4$ abundances in the sample are comparable, they can be easily separated and the final amount of recovered krypton is determined from the area of the Kr peak. Large $CH_4$ concentration in the sample gas requires further measures in order to purify krypton. For example, $CH_4$ is oxidized in a CuO-coated furnace at 800°C, followed by the separation of the produced $CO_2$ and $H_2O$. The purity requirements for the Kr-$CH_4$ separation depend on the analytical method: For LLC, the remaining $CH_4$ volume should allow for a p10-p40 counter filling (for example, < 13 mL-STP $CH_4$ for a p40 filling in a 16 mL counter at 2000 mbar operation pressure). In this case the Kr content in the counting gas is measured by mass spectrometry following the decay activity measurement. For ATTA-3, the detector is insensitive to $CH_4$ except for the reduction of counting rate due to gas dilution. Therefore, the $CH_4$ concentration is preferred to be kept below 50%.

### 4.2.2. University of Illinois at Chicago: Kr

Major constituents of atmosphere, $N_2$, $O_2$ and Ar, have much higher vapor pressure than Kr. Consequently, cryogenic distillation enriches the residual phase in Kr. Large-scale cryogenic distillation with refined engineering control has indeed been the most widely applied method of producing pure gas substances from the atmosphere. A simple method of cryogenic Kr enrichment at an easily attainable constant temperature (77 K) was developed at the University of Illinois at Chicago (Yokochi *et al.*, 2008) (Fig. 7). Pure Kr is obtained by subsequent GC separation and Ti-gettering. The system uses a quadrupole mass spectrometer (QMS) to monitor gas eluent composition during separation, which enables (i) small-scale cryogenic distillation in a controlled manner, (ii) gas chromatographic separation of ppm-level Kr from a significantly large quantity of gas (up to a few liters), and brings (iii) the applicability of the method to natural groundwater samples characterized by variable chemical compositions.

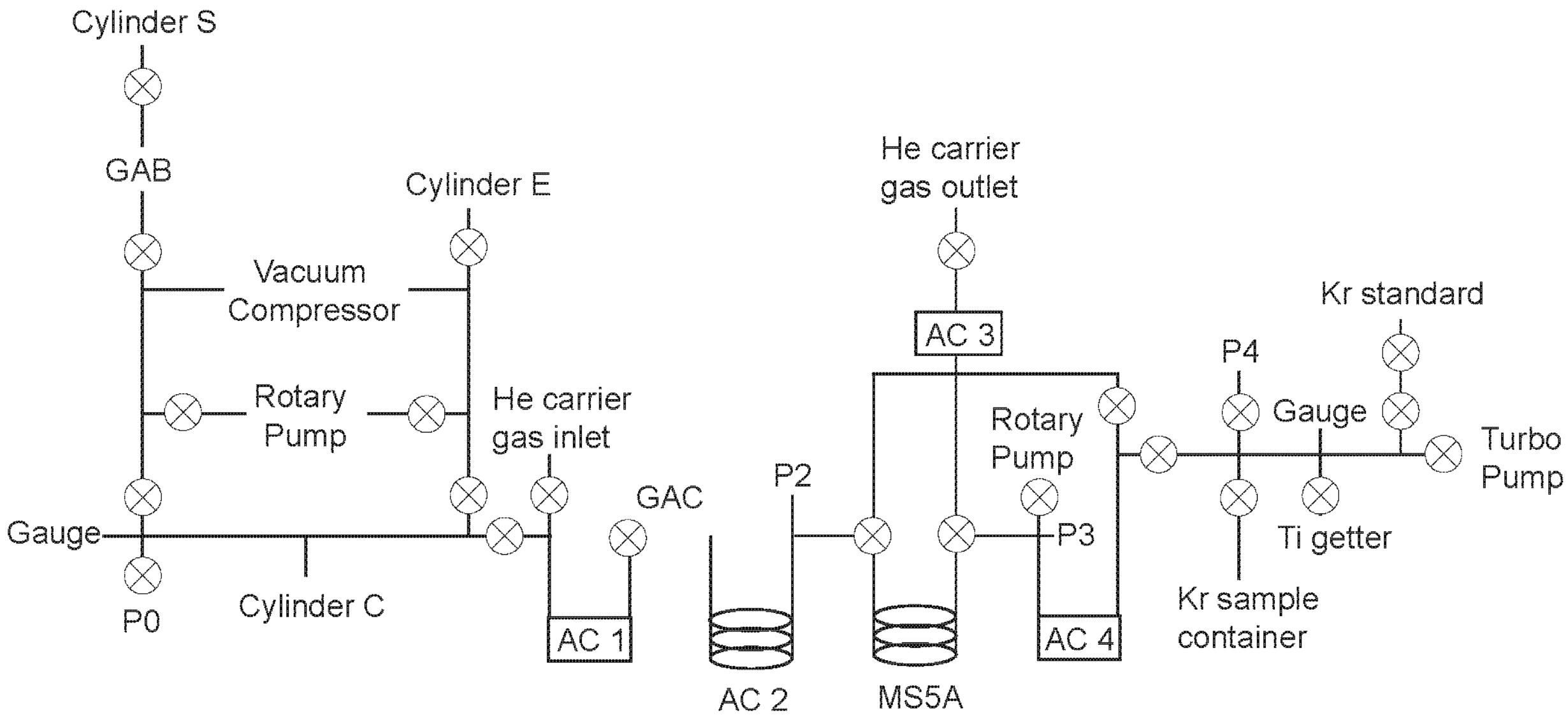

**FIG. 7.** Schematic of the purification system at the University of Illinois at Chicago. Crossed circles represent valves. Acronyms AC, MS, and P represents activated charcoal, molecular sieve, and port, respectively. Figure reproduced from Yokochi *et al.*, 2008.

At first, $H_2O$ and $CO_2$ are removed with a 77 K gas trap at a pressure of < 120 Torr so that Kr does not condense. The remaining sample gas, including Kr, is condensed in an empty, $LN_2$-cooled container at a high pressure (2-3 bars), then distilled at that temperature with the assistance of a vacuum compressor. The rate of the cryogenic distillation is about 3 L $min^{-1}$ for a system dominated by $N_2$, and 0.5 L $min^{-1}$ for a system dominated by $O_2$. The $N_2$/Ar ratio of the distillation gas eluent is continuously monitored to determine the progress of the distillation based on a calculation assuming gas-liquid equilibrium partitioning. Efficient separation of Kr and $CH_4$ from Ar, $O_2$, and $N_2$ can be achieved for up to a few liters of the distillation residue gas by using an activated charcoal column (AC2) at room temperature. The Kr peak following these major species overlaps with the peak tails, but the use of a QMS for monitoring gas composition enables identification and further selective processing of the Kr-enriched portion. At low temperature (around -20ºC), molecular sieve 5-Angstrom separates Kr from $CH_4$ (up to a few hundred $cm^3$ STP) and any remaining $N_2$, $O_2$, or Ar. Pure Kr is collected in a small (0.7 $cm^3$) stainless steel container filled with activated charcoal.

# 5. Applications in Geosciences

## 5.1. Groundwater Residence Time

Over 1.5 billion people depend on groundwater as their primary source of drinking water (Clarke *et al.*, 1996). Increases in world population have led to greater demands on groundwater resources to provide sufficient water for agricultural, industrial, and domestic purposes. Increasing groundwater withdrawals may lead to a higher risk of groundwater pollution from infiltration of inadequately treated wastewaters, as well as reduced groundwater discharge that may affect the amount and quality of surface water flows (Schwartz and Ibaraki, 2011). Global climate change brings the additional threat of diminished groundwater recharge in large regions (Taylor *et al*., 2013). For these reasons it is important to develop a more comprehensive and predictive understanding of the recharge rates, flow paths, and residence times of water in major aquifer systems, to ensure accurate water resource assessments and effective strategies for sustainable withdrawal and protection of water quality (Etcheverry and Perrochet, 2000). Knowledge of recharge rates gives limits on sustainable yields and artificial recharge; knowledge of flow paths can help resolve transnational boundary disputes relating to groundwater production; and knowledge of water residence time distributions, in conjunction with stable isotope compositions, comprises an archive of climatic information that can give insight into long-term aquifer behavior under conditions of changing climate and, when combined with fluid chemistry, provides constraints on the kinetics of water-rock reaction paths.

Here we define groundwater age as the mean subsurface residence time following isolation from the atmosphere, or, more precisely, the soil air. The age can be estimated either from Darcy's Law (based upon hydraulic conductivity, gradient, and porosity) or from measurements of time-dependent abundances of environmental tracers (Maloszewski and Zuber, 1982; Goode, 1996; Varni and Carrera, 1998; Bethke and Johnson, 2008; Torgersen *et al*., 2013). It is one of the most elusive geologic parameters to quantify, despite its crucial significance for water resources, waste management, subsurface reactive transport, and paleoclimate. Groundwater residence times span an enormous range from days to millions of years. However, using numerical simulations to calculate the flow paths and flow rates of groundwater is difficult because of the extreme physical heterogeneity of aquifers (Plummer *et al*., 2013). For example, the permeability of earth materials spans more than 12 orders of magnitude, and the spatial distribution of permeability in aquifers is difficult to characterize. The realistic assessment of water resources and contaminant transport without a valid understanding of flow paths and flow rates is problematic (Sanford, 2011). Environmental tracers that are sensitive to residence times have proven to be effective aquifer evaluation tools, although each sample collected from a well contains information regarding the entire upstream velocity field (Suckow, 2013). Environmental tracers are inherently integrative when compared to aquifer physical property measurements such as water level, permeability, conductivity, and local geologic structure.

A variety of chemical and isotopic residence-time tracers have been used widely for characterizing groundwater systems (Fig. 1). Most of these, including tritium, $^3H/^3He$, chlorofluorocarbons (CFCs), and $SF_6$, are applicable only to young water, i.e., that recharged within the past ~ 60 years (Plummer *et al*., 1993; Cook and Solomon, 1997; Newman *et al*., 2010). Radiocarbon ($^{14}C$) is useful in the range of ~1,000 to 40,000 years, but complications by dissolution of ancient carbonate minerals and biogenic $CO_2$ impart high model-dependence to calculated residence times (Fontes and Garnier, 1979; Plummer and Sprinkle, 2001; Plummer and Glynn, 2013). None of these tracers effectively address time scales in the critical range of ~100 to ~1,000 years, and few tracers are applicable on a time scale of $10^4$-$10^6$ years or more ($^{36}Cl$ and $^4He$). The $^{36}Cl$ method is complicated by variations of the initial $^{36}Cl$ activity and by subsurface input of both stable chloride and nucleogenic $^{36}Cl$ (Bentley *et al*., 1986; Phillips, 2000; Park *et al*., 2002; Phillips, 2013). Many groundwater aquifers do not meet the restrictive criteria for application of $^{36}Cl$-dating, e.g. those containing saline waters and brines, and therefore the $^{36}Cl$ tracer cannot be applied for dating such aquifers, although it remains useful for tracing the origin of salinity in such systems (Phillips, 2000). Accumulation of $^4He$ in aquifer water can thus be used to model groundwater residence time, provided accurate assumptions are made about the rate of supply of $^4He$ to the pore water from the solid phase, as well as the rate of gain and/or loss of $^4He$ by advection and diffusion from or to surrounding formations or the atmosphere (Bethke *et al*., 1999; Torgersen and Stute, 2013). The concentration of $^4He$ in large, old aquifers commonly reflects input from external sources (Torgersen and Clarke, 1985; Torgersen and Stute, 2013).

### 5.1.1. Kr and Ar radionuclides in hydrology

***$^{85}$Kr*** – The ability to determine residence times of young, shallow groundwater is particularly important because of the fact that many shallow aquifers have been subject to contamination by hazardous and toxic substances such as solvents, heavy metals, pesticides, herbicides, endocrine disruptors, and radionuclides. Accurate residence time information for shallow groundwater allows better understanding of the rates of natural attenuation of contaminants, as well as being useful for determining liability and implementing appropriate remediation strategies. The most successful methods available for determining residence times of young (<50 yr) groundwater are: tritium-$^{3}$He (Schlosser *et al*., 1988; Schlosser *et al*., 1989), CFCs, $SF_6$ (Busenberg and Plummer, 2000), and $^{85}$Kr (Smethie Jr. *et al*., 1992). These methods have been compared in detailed studies at a site on the Delmarva Peninsula in eastern Maryland (Ekwurzel *et al*., 1994). At this site, residence times estimated from these tracers all agreed within about 2 years, owing to a combination of high recharge rate, insignificant dispersion, negligible mixing with older water, negligible adsorption-desorption or biodegradation processes, and minimal gas loss to the atmosphere.

Although several groundwater dating methods exist in the 0-50 year age range, all methods have limitations and different methods may fail in a given environment for various reasons such as excess air, degassing, biodegradation, and local contamination. In addition, it is recognized that samples from wells, springs, and streams can be mixtures of water from multiple sources, flow paths, and travel times, and that apparent ages derived from individual dating methods can be misleading in some cases (Boehlke and Michel, 2009; Green *et al*., 2010; Eberts *et al*., 2012). Data from multiple tracers are needed to evaluate problems with individual methods and to resolve age distributions of mixed samples and permit accurate modeling of water movement and chemical transport. $^{85}$Kr can be an especially useful tracer for such purposes because it is relatively immune to some of the potential problems and its input history contrasts with those of some other tracers (Loosli and Oeschger, 1979; Rozanski and Florkowski, 1979; Loosli, 1992; Ekwurzel *et al*., 1994; Althaus *et al*., 2009).

The application of the ATTA (and LLC) method to measuring $^{85}$Kr in young groundwater has several advantages: (1) the decline of CFC input to the atmosphere and the decay of the bomb-produced tritium pulse have begun to reduce the age resolution of these methods; (2) Kr is unaffected by redox reactions such as those involved in biodegradation of CFCs; (3) there is a smaller likelihood of local point sources of $^{85}$Kr than is the case for CFCs and $SF_6$ (e.g. in landfills); and (4) ATTA (and LLC) measures the ratio of $^{85}$Kr/Kr, which is relatively insensitive to gas loss, in contrast to the tritium-helium, CFC, or $SF_6$ methods. In addition, the $^{85}$Kr/Kr ratio is not sensitive to recharge temperature or elevation, or to the presence of excess air.

***$^{39}$Ar*** – $^{39}$Ar has been shown to be useful as a tracer for groundwater residence times in the range of 50-1,000 years (Loosli *et al*., 1970; Loosli and Oeschger, 1980; Loosli, 1983; Loosli *et al*., 1989). $^{39}$Ar produced by the nuclear reaction $^{39}$K(n, p)$^{39}$Ar may be a significant input source in case of high neutron flux and K concentration (Andrews *et al*., 1989; Lehmann *et al*., 1993; Purtschert and Althaus, 2012). In some studies of $^{39}$Ar in groundwater, an isotopic abundance of $^{39}$Ar in excess of the atmospheric abundance has been observed, indicating that nucleogenic $^{39}$Ar from the aquifer rock is transferred to the fluid phase (Loosli, 1983; Andrews *et al*., 1989). This excess $^{39}$Ar is usually considered as an obstacle to determining groundwater residence times (Corcho *et al*., 2007), but may be useful in combination with radiogenic $^{40}$Ar as an indicator of groundwater residence times over a much wider range than that limited by simple decay of cosmogenic $^{39}$Ar (Yokochi *et al*., 2012).

***$^{81}$Kr*** – Some major aquifer systems are dominated by water that has no measurable $^{14}$C, implying residence times exceeding ~40,000 years. This pertains especially to the deeper, confined portions of these aquifer systems. Examples are found in the Great Artesian Basin of Australia, the Nubian Aquifer of Egypt-Libya-Chad-Sudan, and the Guarani Aquifer of Brazil-Argentina-Paraguay-Uruguay. Without knowledge of groundwater ages, it is difficult to construct and validate numerical hydrodynamic models for developing groundwater management strategies that will ensure the optimal use of these critically important water resources by future generations (Sanford, 2011; Schwartz and Ibaraki, 2011). ATTA measurements of $^{81}$Kr enable unprecedented insights into residence times of old groundwater, providing crucial validation to three-dimensional, basin-wide hydrodynamic models. The first application of ATTA (using ATTA-2) to groundwater hydrology determined residence times of old groundwater in the Nubian Aquifer located underneath the Sahara Desert in Western Egypt (Fig. 8)

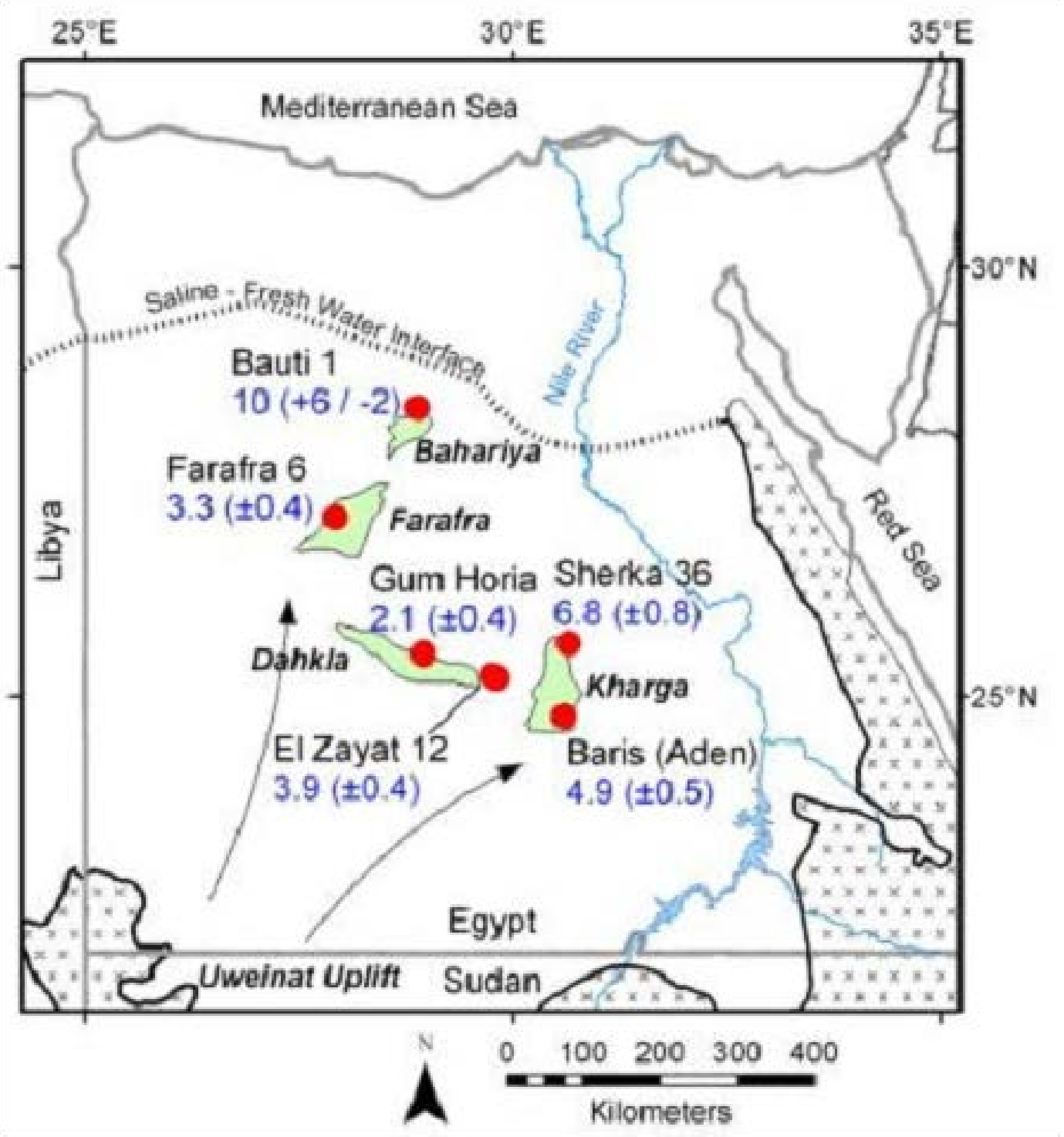

**FIG. 8.** Map showing sample locations (red circles) and their apparent $^{81}$Kr ages (in units of $10^5$ years) in relation to oasis areas (shaded green), Precambrian basement outcrops (patterned), and other regional features. Groundwater flow in Nubian Aquifer is toward the northeast from a recharge area near the Oweinat Uplift in SW Egypt. Figure reproduced from Sturchio *et al*., 2004.

(Sturchio *et al*., 2004). The results of this study revealed the groundwater age and hydrologic behavior of this huge aquifer, with important implications for climate history and water resource management in the region. Also demonstrated was the consistency between $^{81}$Kr and $^{36}$Cl measurements, and there were near-blank activities of $^{85}$Kr, $^{39}$Ar, and $^{14}$C in these waters as expected. Additional studies of old aquifers using $^{81}$Kr are now in progress (e.g., the Great Artesian Basin in Australia and the Guarani Aquifer in South America). Recent work on South African deep mine waters by Purtschert *et al*. (2013b) has revealed appreciable *in situ* production of $^{81}$Kr in this rock-dominated system of fractured Archean basement.

### 5.1.2. Recommendations for future work

Hydrologists at the TANGR 2012 workshop noted the following in terms of both the importance and the implementation of $^{39}$Ar and $^{81}$Kr in critical groundwater research:

- Sample size is an important issue for groundwater studies, especially in low permeability formations. Ultimately, sample volumes of 1 L or less would be desirable; however, *many* important regional aquifers yield large amounts of water and, provided that efficient systems are employed for gas extraction in the field, the present sample size of about 100-200 L is workable. Implementation of $^{81}$Kr and $^{39}$Ar to these regional aquifer systems should start immediately, while analytical advances that reduce sample size continue in order to expand the utility of these tracers.

- Data sets for groundwater studies are inherently small due to the high cost of installing monitoring wells. This makes the use of integrative tracers such as $^{81}$Kr and $^{39}$Ar even more important. A limited number of shallow aquifer studies at intensively-monitored sites have been conducted, but there is a need for such campaigns at larger (10 - $10^2$ km) scales.

- Many groundwater systems are rapidly changing due to recent pumping. Recharge rates that led to the current distributions of tracer concentrations may no longer exist. It is critical that integrative tracers such as $^{81}$Kr and $^{39}$Ar be measured in systems before additional (and sometimes massive) pumping occurs in order to effectively utilize these tracers for evaluating the hydraulic properties of the aquifer. This fact makes the implementation of residence time tracers urgent to accomplish before critical information is lost.

- The utility of existing tracers such as $^{14}$C and $^{4}$He could be greatly enhanced by calibrating them against geochemically simple tracers such as $^{81}$Kr and $^{39}$Ar. Such calibrations could greatly enhance existing data sets along with new data collection efforts.

- The nature of groundwater transit times in samples collected from wells and discharge locations such as seeps and springs is such that a typical sample may contain a large range of groundwater age values. This makes the use of multiple tracers that speak to different transit times important (Zuber, 1986; Reilly *et al*., 1994; Corcho *et al*., 2007; Zuber *et al*., 2010). Each application of tracers for longer timescales (e.g., $^{14}C$, $^{36}Cl$ or $^{81}Kr$) needs the information from the shorter-timescale tracers ($^{3}H/^{3}He$, $^{85}Kr$, $^{39}Ar$) to allow deconvolution of the age distribution in the sample and to identify possible admixtures of young water. Measurement of integrative tracers with relevant half-lives is a valuable strategy for understanding groundwater flow systems and for planning further detailed investigations.

## 5.2. Ocean Ventilation

Ventilation is the primary conduit for passing signals from the atmosphere and the climatic system to the interior of the ocean. For instance ventilation directly influences the distributions and controls of natural and anthropogenic carbon (both organic and inorganic), which in turn directly influences the $CO_2$ concentration in the atmosphere, and hence the radiative forcing of the earth's climate system (Sabine and Tanhua, 2010). The $CO_2$ invasion into the interior ocean also influences the pH of the ocean and the saturation state of aragonite and calcite, two important minerals for shell building marine organisms (Feely *et al*., 2004; Orr *et al*., 2005). Ventilation of the ocean transports heat from the atmosphere to the interior ocean where the high heat capacity of the ocean, as compared to the atmosphere, is important for the climate (Purkey and Johnson, 2010). Similarly, ventilation processes are responsible for transporting oxygen to the interior ocean from the surface, where gas exchange with the atmosphere and photosynthesis drive an oxygen flux to the ocean (Keeling *et al*., 2010; Stendardo and Gruber, 2012). The oxygen concentration in the ocean is important for marine life and for several biogeochemical processes, such as de-nitrification, a process that removes bioavailable nitrogen from the ocean in low oxygen environments. Changes in ocean ventilation will thus not only directly impact the global climate but also have consequences for marine biology, and in the end will impact society through factors such as sea-level rise, climate and fishery sustainability.

Ventilation rates of the ocean can be determined by measurements of tracers (e.g. Broecker and Peng, 1982; Bender, 1990; Waugh *et al*., 2003; Fine, 2011). For instance, oxygen levels have been used for this purpose, although that involves assuming a value for oxygen consumption rates (Karstensen and Tomczak, 1998). More accurate estimates can be made with tracers that are conservative in the ocean. The most commonly used tracers are tritium-$^{3}He$ (e.g. Rooth and Östlund, 1972; Jenkins and Clarke, 1976), the chlorofluorocarbons (CFCs) that were introduced since the 1940's to the atmosphere (e.g. Gammon *et al*., 1982; Weiss *et al*., 1985), or $SF_6$ that has been entering the atmosphere since the early 1960's (e.g. Law *et al*., 1994; Law and Watson, 2001). For waters with slower ventilation the most commonly used tracer is radiocarbon ($^{14}C$) (e.g. Revelle and Suess, 1957; Broecker *et al*., 1960; Broecker *et al*., 1985). The $\Delta^{14}C$ age has some systematic uncertainty mainly due to the long equilibrium times with the oceanic carbonate system in the surface mixed layer (on the order of 10 years), i.e. newly ventilated surface waters initially have a non-zero age on the "$\Delta^{14}C$–clock" (Broecker and Peng, 1982; Stuvier *et al*., 1983; Matsumoto, 2007). There are also small uncertainties associated with downward flux of organic particles that dissolve in the deep ocean. By accounting for the "reservoir-age" of the surface mixed layer, Matsumoto (2007) estimated the ventilation ages of the deep-ocean to be less than 1000 years in most areas except the North Pacific.

$^{39}Ar$ observations would fill a temporal gap in ocean tracer observations. Large areas of the ocean, particularly the deep Indian and Pacific Oceans, are filled with "old" water that does not contain CFCs or other anthropogenic tracers. Argon equilibrates with the atmosphere rapidly (on the order of days to weeks) so that most of the ocean surface can be considered as saturated with respect to the $^{39}Ar/Ar$ ratio, with the exception of the Southern Ocean (Schlosser *et al*., 1994a) and Arctic Ocean (Schlosser *et al*., 1994b) where deep water can be formed under the ice. The well-known input function at the surface for most of the ocean, the chemical inertness of argon, and the half-life of 269 years combine to make $^{39}Ar$ ideal for deep ocean ventilation studies.

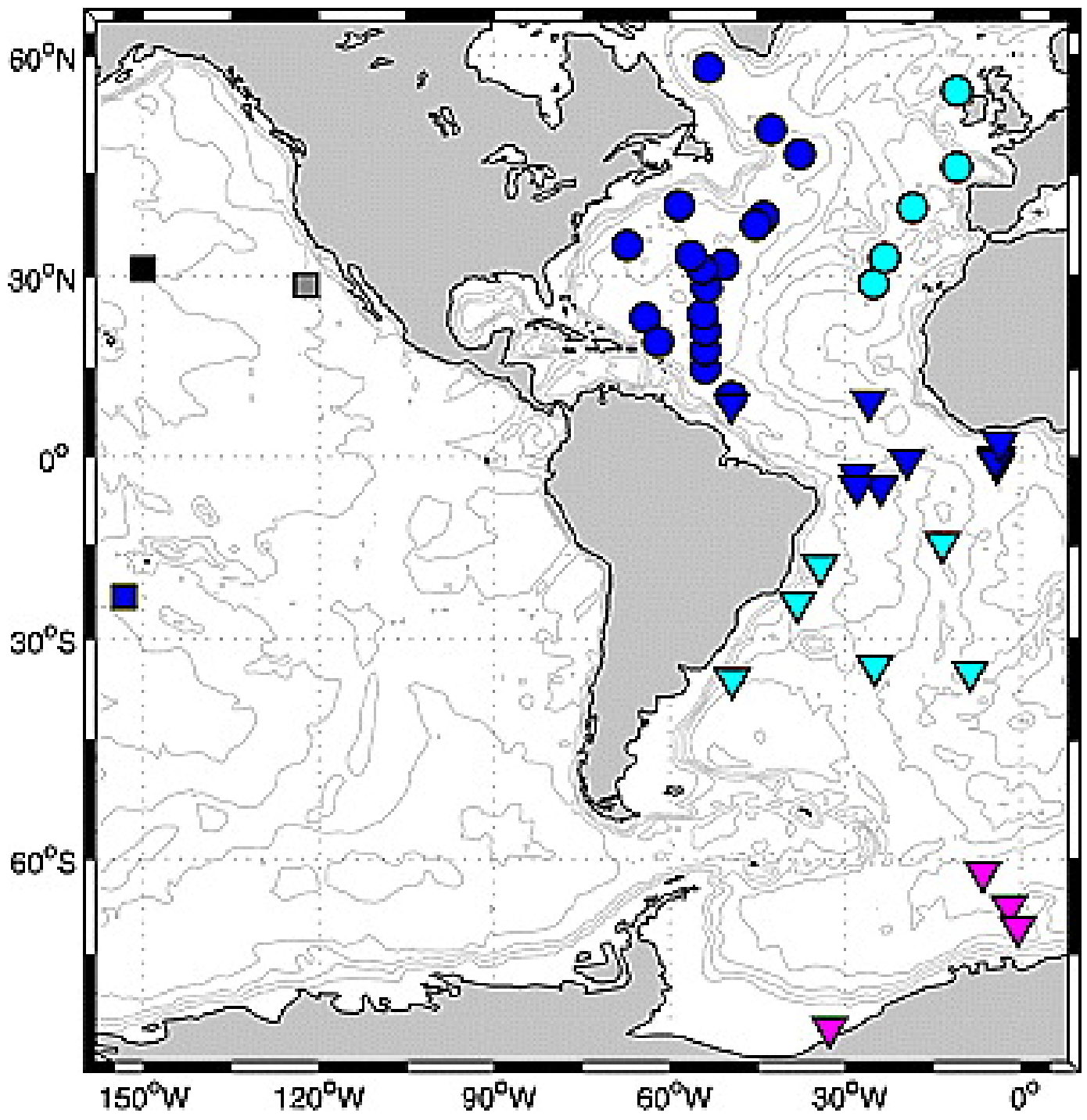

**FIG. 9.** Locations of interior ocean $^{39}Ar$ measurements. Figure reproduced from Holzer and Primeau, 2010.

A limited number (~ 125) of interior ocean $^{39}Ar$ data are available from large volume sampling and decay counting; most (but not all) of the sample locations are shown in Fig. 9. Two of the few available depth profiles from the Pacific Ocean, where the world's oldest oceanic waters reside, are shown in Fig. 10. The profiles show decreasing $^{39}Ar$ concentration (increasing age) through the water column to a minimum $^{39}Ar$ concentration at intermediate depths, and then increasing concentration towards the bottom due to the influence of younger Antarctic Bottom Water (AABW). AABW forms near the surface around Antarctica, spreads northward along the bottom and upwells to intermediate depths. This flow pattern is reflected in the higher $^{39}Ar$ concentration at 23°S than at 31°N, and in the lowest $^{39}Ar$ concentration at intermediate depths. The $^{39}Ar$ age of this intermediate water is about 900 years. In a separate work, $^{39}Ar$ concentrations of up to 57% modern have been reported for waters in the Atlantic sector of the Southern Ocean (Weddell Sea) (Schlosser *et al.*, 1994a). These measurements directly determined the initial $^{39}Ar$ concentrations of newly formed bottom water in the Southern Ocean and serve as one reference point for age estimates in the ocean interior using $^{39}Ar$. A basin-scale study of $^{39}Ar$ and $^{14}C$ has been reported for the Northeast Atlantic (Schlitzer *et al.*, 1985). It revealed consistent results between these isotopes with respect to mean residence times of the deep waters in this ocean basin.

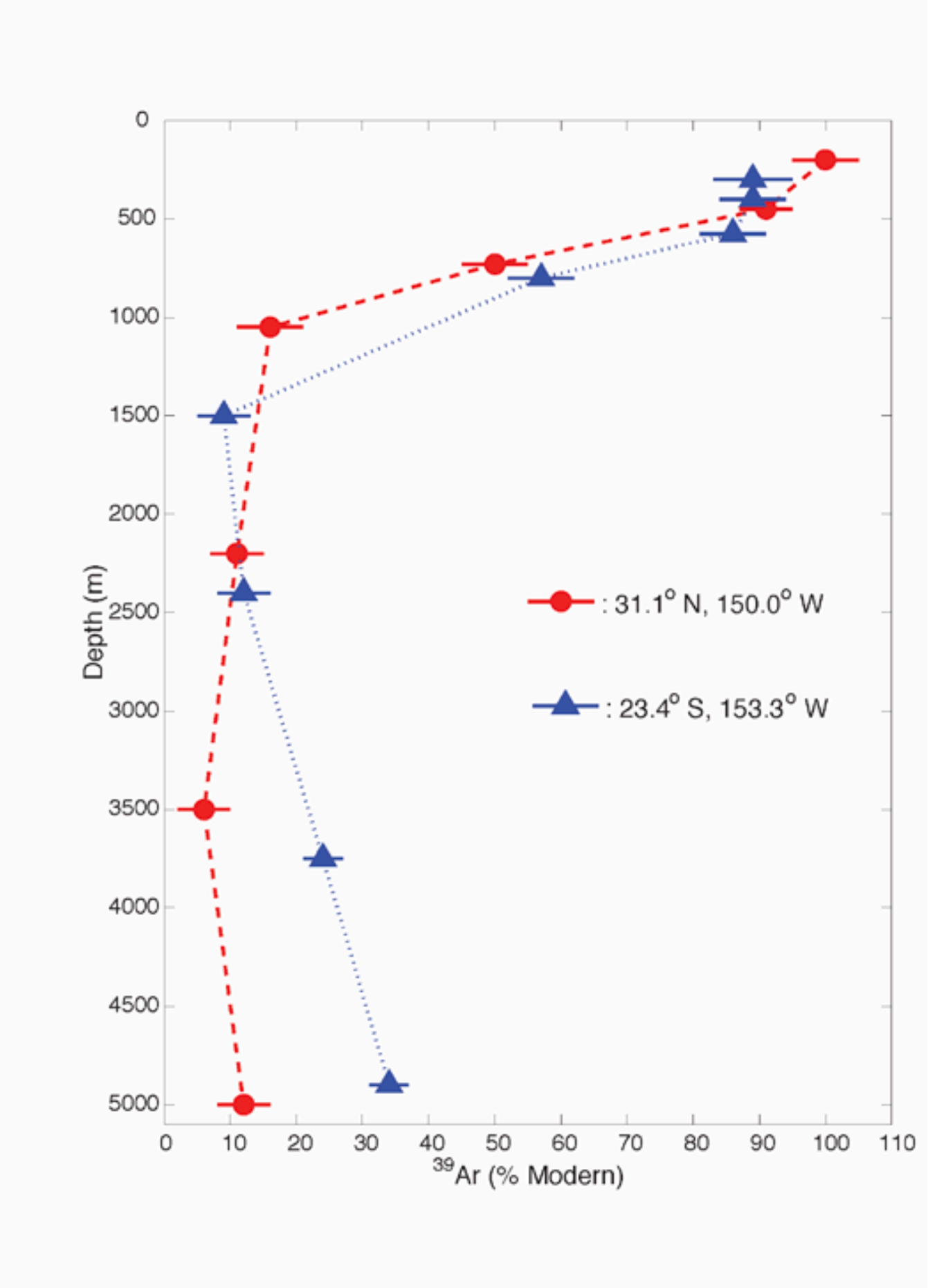

**FIG. 10.** $^{39}Ar$ profiles from the Pacific Ocean (see Figure 9 for station locations) sampled during the GEOSECS program in the 1970s (Craig and Turekian, 1976). See Rodriguez (1993) and Loosli (1989) for data and additional information.

Holzer and Primeau (2010) used these measurements in a formal study to show that the $^{39}Ar$ data could reduce the uncertainty in ocean ventilation (represented as a distribution of water parcel transit times from their origins at the ocean surface) by, on average, about 50% as compared to a case where only CFCs and $^{14}C$ were considered. The reduction in uncertainty is particularly significant for water with long ventilation times, such as in the deep Pacific Ocean where the CFCs have not penetrated, as well as regions with low CFC concentrations such as the deep thermocline where a large amount of older CFC-free water has mixed with a small amount of younger CFC bearing water. They conclude their study by stating that *"It is no question that an effort to produce a globally gridded $^{39}Ar$ data set would greatly help reduce the uncertainty in our knowledge of oceanic transport and ventilation, which in turn would greatly benefit rigorous estimates of the ocean's ability to take up and sequester anthropogenic carbon."*

Similarly, a study by Broecker and Peng (2000), which compared $^{39}$Ar and $^{14}$C ages of the deep ocean, concluded that *"It is clear that a more dense survey of $^{39}$Ar with higher accuracy measurements would prove of great value in constraining ocean general circulation models."* Taken together, $^{39}$Ar and $^{14}$C provide strong constraints on the ratio of mixing to advection in the deep ocean.

Opportunities for future measurements of $^{39}$Ar include a suite of about 1000 samples from the Atlantic Ocean collected in the 1980s. These samples have been measured for $^{14}$C, are available and would constitute an excellent baseline for characterizing Atlantic deep ocean ventilation about 30 years ago. The geographic coverage of these samples is remarkably good (Fig. 11). For future sampling there are ample opportunities to collect samples on, for instance, GO-SHIP repeat hydrography (http://www.go-ship.org) that repeats oceanographic sections on a 10 – 15 year time scale to document changes in ocean properties, and / or during cruises within the GEOTRACES project (http://www.geotraces.org/) that will be mapping the global ocean distributions of various trace elements and isotopes over the next decade.

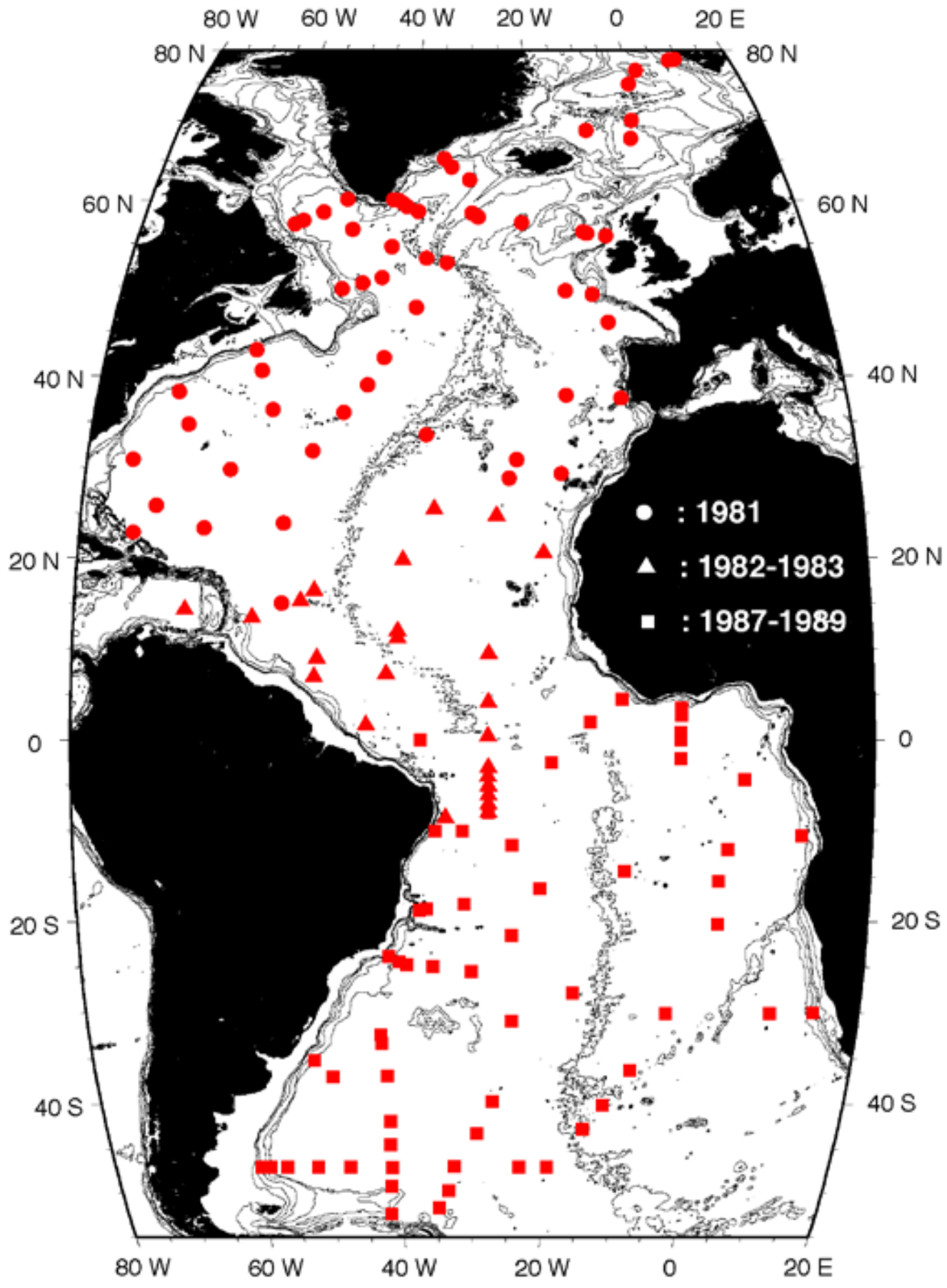

**FIG. 11.** Locations of oceanographic stations where large volume (250 liters) samples were collected for measurements of $^{14}$C and $^{85}$Kr during the Transient Tracers in the Ocean (Brewer *et al.*, 1985) and South Atlantic Ventilation Experiment programs. Either the argon fractions or the entire extracted gas samples have been archived at Lamont-Doherty Earth Observatory.

The Pacific Ocean (3 stations) and the Indian Ocean (no samples) are heavily under-sampled with respect to $^{39}$Ar. These two ocean basins have the oldest deep water, i.e. $^{39}$Ar data would be particularly useful for these basins. Conducting repeat $^{39}$Ar measurements in the Atlantic Ocean for a comparison of ventilation ages between the early 1980's and the 2010's would also be interesting.

Sampling of roughly 1000 samples in the Pacific Ocean, 500 in the Indian Ocean, 200 in the Arctic Ocean and 300 repeat samples in the Atlantic Ocean would provide a reasonable one-time coverage. This would mean ~2000 new samples and measurements of 1000 archived samples. There are two important considerations for future measurements and sampling. The desirable measurement precision should provide an age resolution of about ± 25 years, comparable to the age resolution of $^{14}$C measurements, which ranges from ± 15 to ± 30 years. Ideally the water sample volume should be 1 L or less so that the samples can be drawn in parallel to other samples. Sample sizes of 10 L using the standard 10-L sample bottles for collection and 30 liters using the 30-L Niskin bottles would also be feasible.

## 5.3. Ancient Glacial Ice

Polar ice cores have been used to reconstruct Earth's past climate and atmospheric composition as far back as 800,000 years (Jouzel *et al.*, 2007). A range of techniques is used for the dating of ice cores, including annual layer counting, ice flow modeling, age marker synchronization and orbital tuning. Radiometric dating tools for old ice are currently not widely used because of several distinct drawbacks. Radiocarbon dating on the $CO_2$ present in air bubbles is complicated by *in situ* cosmogenic $^{14}$C production in the ice (Lal *et al.*, 1990). Other radiometric methods with limited precision and/or resolution rely on the incidental inclusion of meteorites, tephra layers or organic impurities (Jenk *et al.*, 2007; Dunbar *et al.*, 2008).

$^{81}$Kr could potentially be used for dating of old ice with ages ranging from 100 kyr - 1500 kyr (Oeschger, 1987). $^{81}$Kr dating has several advantages: (1) the technique is widely applicable as all polar ice contains air bubbles; (2) the technique does not require a continuous or undisturbed ice stratigraphy; (3) there is no *in situ* $^{81}$Kr production, as is the case for $^{14}$C; (4) the dating method gives an absolute age estimate. The main disadvantage is the large sample size (>40 kg), which has so far precluded its use in ice core studies. Unlike groundwater, there is no need to account for age distribution changes due to stratigraphic disturbance. The typical pattern of stratigraphic disturbance in ice cores is one of discrete packages of intact stratigraphy, on scales of many meters, separated from neighboring packages by faults or zones of concentrated shear. Thus within any particular meter of ice core, there is almost always a locally-intact piece of ice (see, for example, Dahl-Jensen *et al.*, 2013).

Old ice can be obtained not only from deep ice cores, but also at ice margins and Antarctic blue ice areas (BIAs) where it is being re-exposed by ablation (Sinisalo and Moore, 2010). For paleoclimate studies this provides an interesting alternative to ice coring, as large volume samples can easily be obtained. During the last few years the ice ablating at the Taylor Glacier BIA (Fig. 12) has been dated using stratigraphic matching techniques (Petrenko *et al*., 2006). An experiment is being carried out to use the well-dated stratigraphy of Taylor Glacier to test the feasibility and accuracy of $^{81}$Kr-dating of old ice for the first time (Buizert *et al*., 2013). If successful, the experiment would demonstrate that $^{81}$Kr-dating is indeed feasible. This could pave the way for reliable $^{81}$Kr-dating of other BIAs, which would significantly enhance the scientific value of BIAs in Antarctica. As the required sample size continues to decrease, $^{81}$Kr dating of ice core samples might be possible in the future.

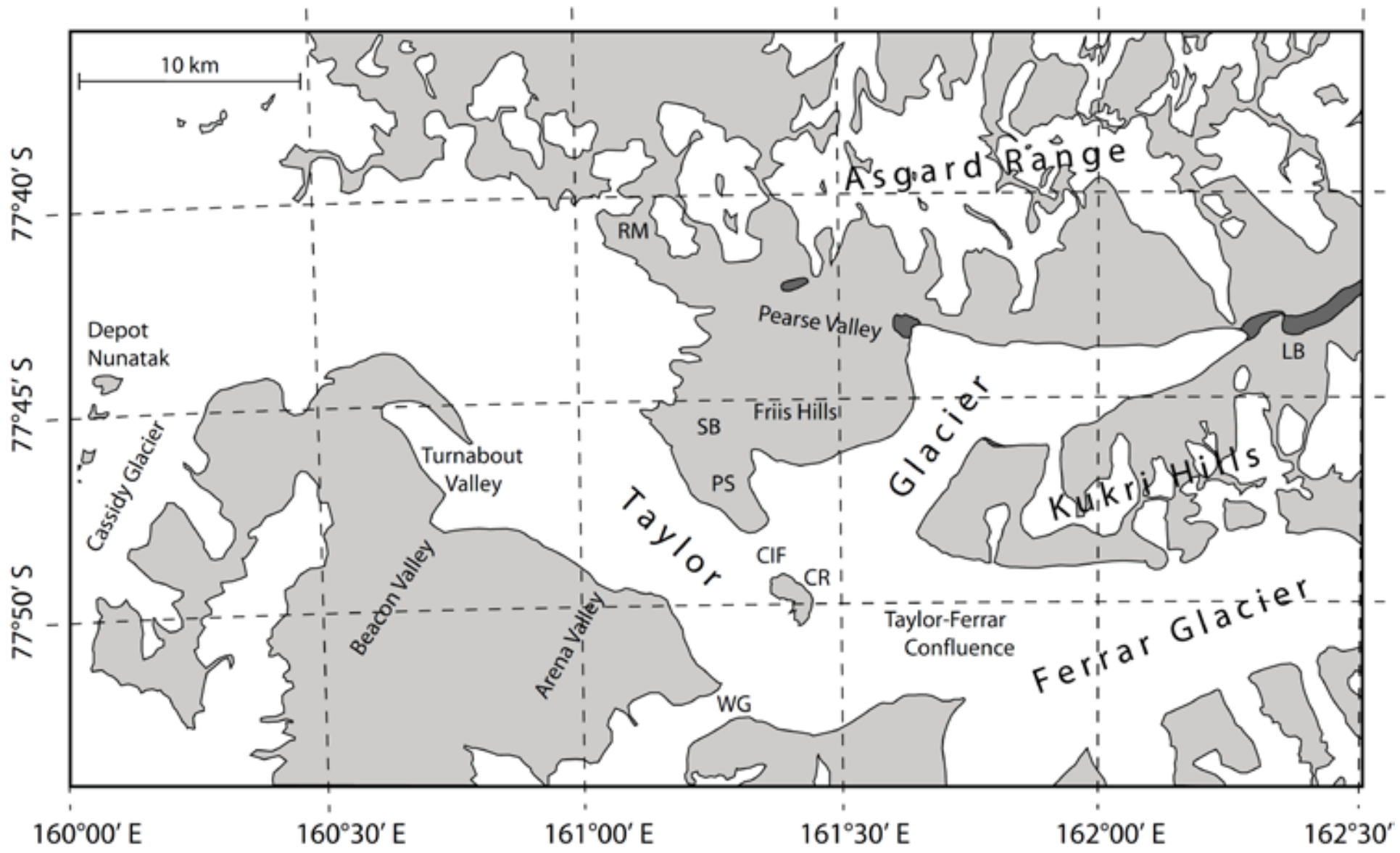

**FIG. 12.** Taylor Glacier is an outlet glacier of the East Antarctic ice sheet that originates at Taylor Dome, and terminates in Taylor valley of the McMurdo Dry Valleys, Victoria Land. The map shows the lower reaches of the glacier where sublimation-driven ablation exceeds snow accumulation, giving rise to a blue ice area (BIA). Figure reproduced from Kavanaugh *et al*., 2009.

A major thrust of current ice core research is to extend the existing 800,000 yr record back in time to 1.5 million years. Such a record would permit a test of the hypothesis that falling atmospheric $CO_2$ concentration caused the shift in the pace of ice ages from a regular 41,000 yr period (which prevailed during the 1.5-1.2 Ma interval) to the current irregular ~100,000 yr period. A second goal is to understand the puzzling absence of the 23,000-yr orbital precession period during this time of 41,000-yr cycles in the marine isotope record of ice volume, despite its prevalence during the past million years (Severinghaus *et al*., 2010). An international planning group known as the International Partnerships in Ice Core Sciences (IPICS) has formed around this challenge, known as the Oldest Ice quest, because the difficulties are formidable and are thought to be beyond what any one nation can achieve (http://www.pages-igbp.org/ipics/).

Blue Ice Areas (BIAs) hold promise as potential sources of 1.5 Myr ice, because very old ice is known to stagnate in cold, high-altitude BIAs (Spaulding *et al*., 2012). The problem is that there are literally thousands of candidate BIAs in Antarctica, and the dating of these by conventional methods that involve hundreds of samples in an age transect is logistically prohibitive. $^{81}$Kr dating offers a way around this obstacle because a single sample of ice can quickly establish a ball-park age in a rapid-reconnaissance style survey. In this way the large number of BIAs can be narrowed to two or three candidates for intensive study using the more precise (and labor-intensive) dating techniques. Such a dating campaign would also rapidly increase the available knowledge base for calibration of ice flow models, which are being used to predict Antarctic ice sheet evolution and sea level rise.

Another distinct advantage of $^{81}$Kr dating is that it can be used in situations where ice is stratigraphically disturbed, and conventional techniques fail due to the fact that they require stratigraphic continuity (for example, to match fluctuations in oxygen isotopes to the marine isotope record). This is because a stand-alone $^{81}$Kr date can be obtained on a single piece of ice. Very old ice in BIAs or ice cores is highly likely to be stratigraphically disturbed, owing to the fact that it has travelled near the bedrock where shear has induced folding and faulting. For example, the recent recovery of ice from the last interglacial period in Greenland, a climatic period similar to what is expected in the coming century, required reconstruction of the climate record from five discontinuous folded and faulted pieces of ice (NEEM Community Members, 2013).

For these reasons, the continuing development of increased precision and reduced sample size in $^{81}$Kr dating is of great interest to the glaciological and ice core community. For BIA related research in particular, the current sample size of 50 kg is already practical using technology developed for the Taylor Glacier project (Fig. 13).

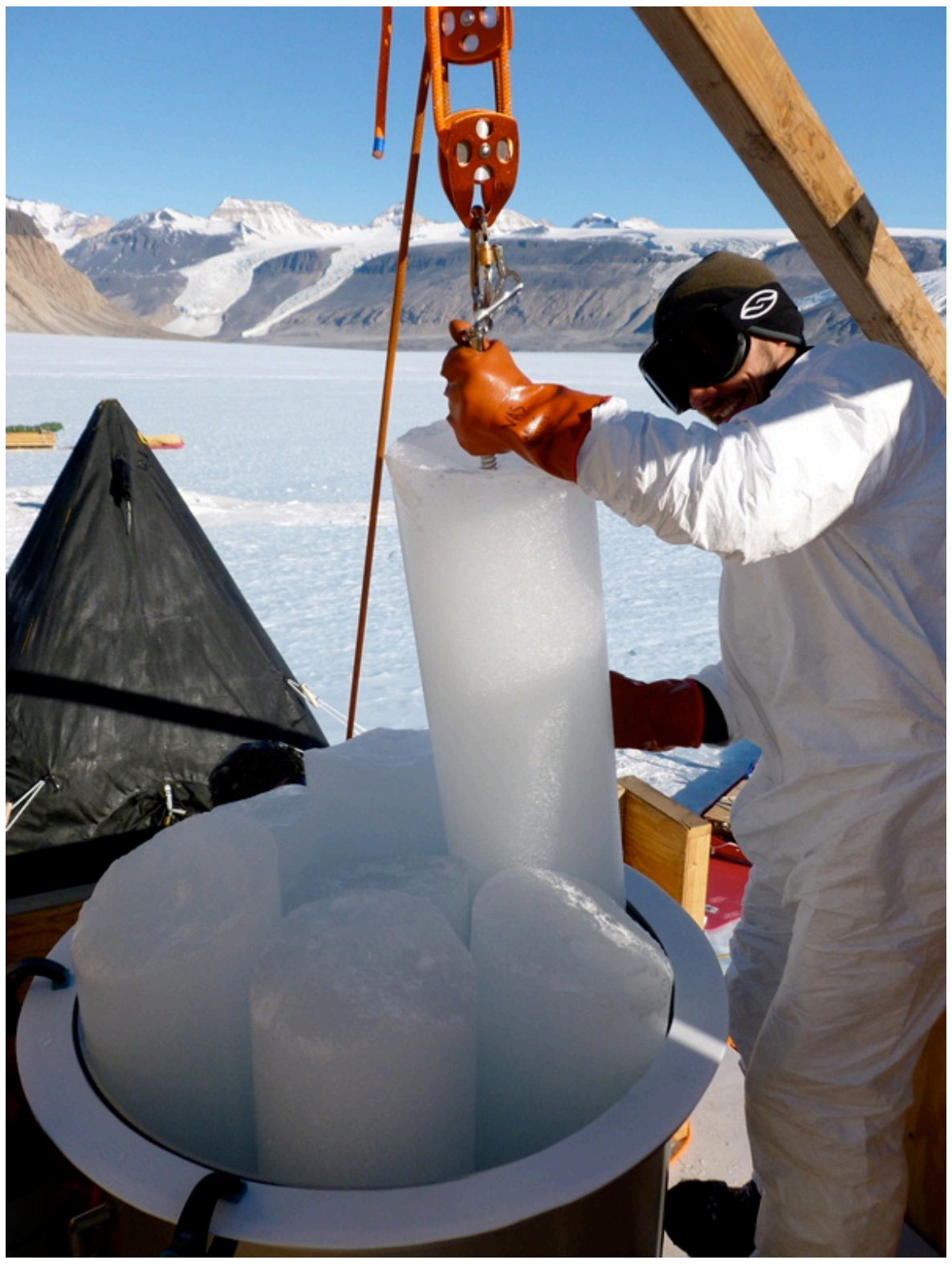

**FIG. 13.** Large-volume ice sampling for $^{81}$Kr dating at Taylor Glacier. The ice cores are 24 cm in diameter and the mass of ice being loaded into the vacuum extraction vessel is ~ 300 kg. Photo credit: V.V. Petrenko.

## 5.4. Deep-Seated Fluids

The Earth's crust is a vast reservoir of resources formed via natural geological processes driven by crustal fluids and fluid flow (Fig. 14). Crustal fluids mediate chemical reactions, and transport, concentrate or disperse elements in the crust. In this context, determining the origin of crustal fluids, the time scales of fluid transport and residence are essential for understanding the geochemical cycle of elements, as well as risk and resource management. For instance, although interactions between groundwater and gas, oil and hydrothermal fluids have been well studied, the fluid migration rate or storage time in crustal reservoirs has rarely been determined due to the lack of optimal tracers. Volcanic gases play the role of geochemical messengers by transporting volatiles from the deep Earth. A long-standing problem in the study of volcanic gases is that the modern atmospheric and meteoric component cannot be distinguished from the ancient subducted atmospheric component. The noble gas radionuclides can be used to help address this problem, leading to a more accurate chemical and isotopic characterization of the underlying mantle. This, in turn, sets significant constraints on the concentrations of volatile elements in the mantle, the origin of the terrestrial atmosphere, and geochemical cycles of volatile elements through long-term geological processes. The ability to routinely analyze fluid samples for noble gas radionuclides ($^{39}$Ar, $^{81}$Kr and $^{85}$Kr) will provide an indispensable tool for deriving a thorough understanding of the origin, evolution and migration of deep-seated fluids. Two relevant examples are provided below for volcanic and natural hydrothermal systems.

### 5.4.1. Geological water cycle

Fluids discharging from hydrothermal and volcanic systems through fumaroles, hot springs and geothermal wells are dominated by water (> 90%), $CO_2$, sulfur ($SO_2$ and $H_2S$), and chlorine. In addition to these major components, $N_2$, $H_2$, $CH_4$, CO, HF and the noble gases generally contribute to the gas composition. Work over the past two decades has utilized the technique of combining helium ($^3$He and $^4$He) with other stable isotope systems to estimate the relative contributions of C and N from the mantle, the subducting slab, and atmospheric and crustal sources (see reviews by Hilton *et al*., 2002; Sano and Fischer, 2013). This work has provided insights into the recycling of volatiles from the Earth's surface to the interior through subduction with implications for the evolution of mantle volatile compositions and the exchange between deep earth and surface reservoirs.

While the sources of C and N in these systems are relatively well constrained in terms of mantle wedge, slab and crust, the ultimate sources of water and sulfur remain somewhat more enigmatic. In addition and perhaps more significantly, the timescales of volatile exchange between the Earth's interior and the atmosphere is currently mainly constrained by models involving stable isotopes (Marty *et al*., 1989; Marty and Tolstikhin, 1998; Marty and Dauphas, 2003). In the mantle, water is the most significant volatile species due to its abundance and effect on melting and element transport processes (Wallace, 2005). Slab derived water lowers the melting point of mantle peridotite which ultimately results in water-rich and highly explosive arc magmas (Tatsumi, 1989). On the other hand, surface derived water (i.e. meteoric water) is the driving force for surficial water-magma interactions resulting in highly explosive phreatic eruptions. Meteoric water is also the agent for heat transfer from the magma to the surface in volcanic hydrothermal and geothermal systems. Therefore, understanding the source and relative proportions of magmatic (mantle derived) versus meteoric water is critical for evaluating the global deep water cycle as well as investigating the sources of heat in geothermal systems.

Global data suggest the existence of an end-member in $\delta^{18}O$ and $\delta D$ space that is related to water derived from the devolatilization of the subducting slab. In the crust or close to the surface this “arc-type” water mixes with meteoric water resulting in isotopic compositions that lie along mixing lines extending from the meteoric water end-member to the arc-type water end-member (Taran *et al*., 1989; Giggenbach, 1992;). The end-member composition of arc-type water has been confirmed by $\delta D$ measurements of $H_2O$ in melt inclusions from volcanic arcs, supporting the idea of a unique slab-derived water end-member (Shaw *et al*., 2008). Magmatic volatiles often interact with meteoric water (and other volatiles in rocks/soils in the crust) during ascent from depth to the surface where fluid samples are obtained. It is therefore, desirable to distinguish magmatic water, which would have no detectable $^{81}Kr$ from the surface components, including modern air-derived volatiles. Kr- isotope systematics, therefore, would provide an independent method to quantify the meteoric and magmatic component of volcanic gas discharges. Such data would provide better constraints on the global magmatic water cycle.

Application of $^{81}Kr$, $^{85}Kr$, and $^{39}Ar$ measured in volcanic emissions that have stable isotope signatures of slab-derived water, combined with other geochemical tracers (C, N, He isotopes and abundance ratios), would provide unprecedented insights into the ultimate source and age of the fluids transporting these gases. For instance, slab derived water is expected to have an age on the order of 1 Myr (in the range of $^{81}Kr$ dating) and surface (meteoric water) that has interacted with the crust would have expected ages of $10^2$-$10^4$ years (overlapping with the range of $^{39}Ar$ dating). In contrast, shallow and surface derived water is expected to be relatively young ($<$ 100 years and in the range of $^{85}Kr$ dating). The combination of radionuclide isotopic abundances with O and H stable isotopes as well as water fluxes constrained with other methods has the potential of elucidating the extent and rate of water release from the Earth’s interior, transport of water into the deep mantle (re-gassing of the mantle) and evaluating the rates of surface water transport to and from crustal magma reservoirs. Similar investigations at continental rifts could provide complementary insights on the time scales of mantle water degassing (Fig. 14). On a smaller scale, the rate of heat exchange between crustal magma bodies and adjacent geothermal reservoirs could be determined by relationships of heat versus age and source of geothermal water sampled at the surface. This approach could provide valuable insights into the potential longevity of a geothermal system and could be a powerful tool in managing the geothermal resource. This application is further discussed in the next example.

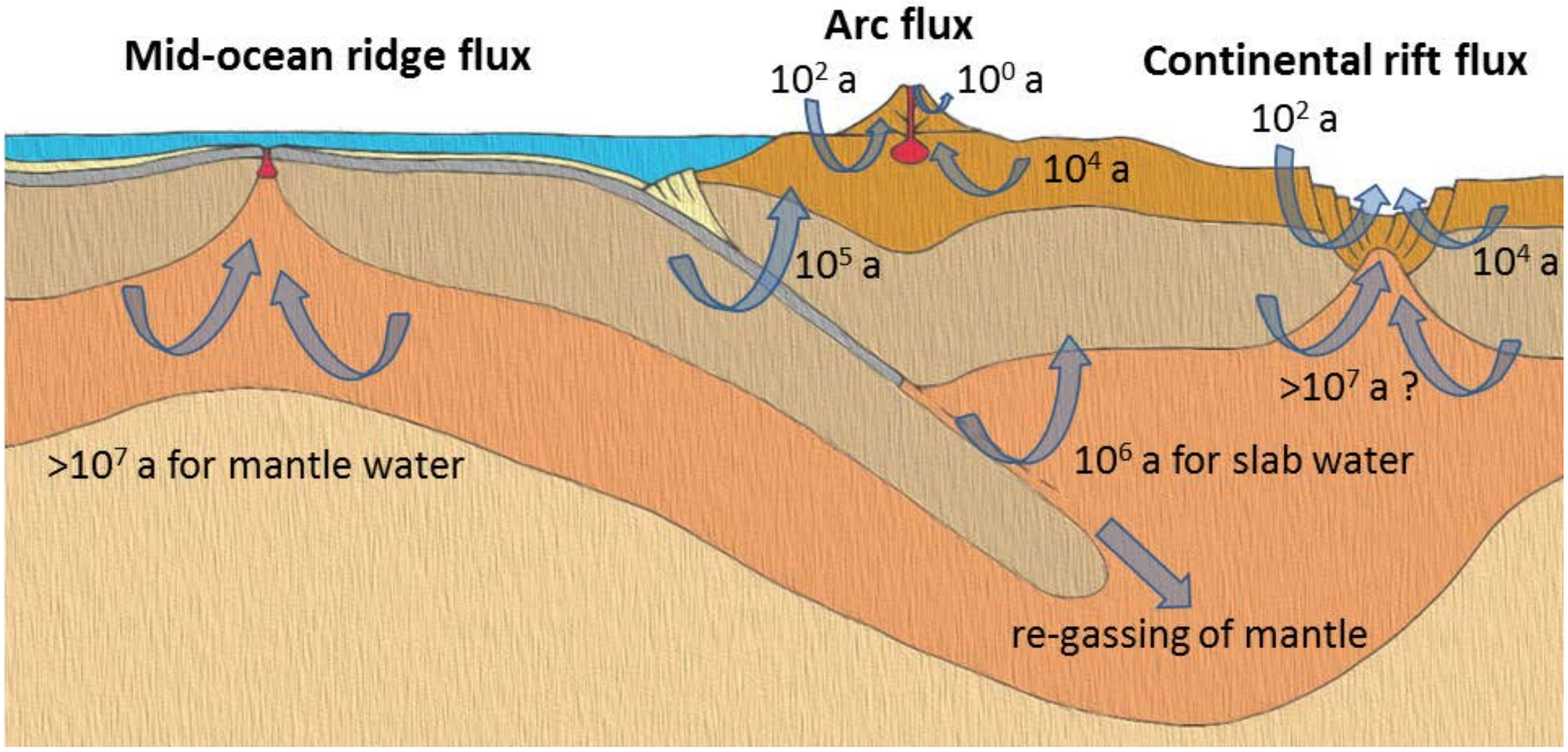

**FIG. 14.** Towards a better understanding of the water cycle and the rates of water exchange between the surface and deep Earth (numbers are estimates of the age of water in various locations). Figure credit: W. Jiang.

### 5.4.2. Fluid residence and migration in the crust

In addition to the atmosphere derived radionuclides ($^{39}Ar$, $^{81}Kr$ and $^{85}Kr$), crustal fluids can acquire stable and radioactive radiogenic noble gases (e.g. $^{4}He$, $^{40}Ar^*$ and $^{39}Ar$) that are continuously produced by the decay of U, Th and $^{40}K$, and by nuclear reactions in the crustal minerals. These “crustal” stable and radioactive nuclides may be released from their production sites and incorporated into the fluid as additional natural spikes to trace fluid flow and residence ages. Their concentrations in a crustal fluid will be a function of the accumulated concentrations and production rates in the reservoir rock and the rate of transfer from the rocks to the fluid phase. In the absence of subsurface production and/or transfer of those isotopes into the fluid phase, the fluid migration time between two sites within a flow path can be determined simply by measuring the abundance of a single atmospheric derived radionuclide with an appropriate half-life. In the case where transfer of subsurface-produced isotopes from the rocks to the fluid phase occurs, the concentration of radiogenic stable isotopes in the fluid phase will increase with time or flow distance in the same manner as radiogenic $^{4}He$ and $^{40}Ar^*$. However, due to uncertainties related to prior accumulation and release rates from crustal minerals, they have not been used routinely as reliable groundwater chronometers. By combining radiogenic $^{40}Ar^*$ with nucleogenic $^{39}Ar$, which acquires a steady-state concentration in minerals after ~1800 years, uncertainties related to poorly constrained release rates can be minimized. Accordingly, the migration time of a fluid between two points within a single flow path can be determined using the ratio of radiogenic $^{40}Ar^*$ and $^{39}Ar$. This methodology can be applied to in-situ produced $^{81}Kr^*$ (Purtschert *et al*., 2013), provided that stable isotopes of Kr or other noble-gas elements are also produced in related processes. As with other systems, $^{85}Kr$ can be used as an unprecedented and unique tracer of atmospheric contamination, distinguishing modern atmospheric contributions from ancient ones.

In crustal fluids, stable noble gases have been used for investigating the processes occurring in subsurface reservoirs, during (i) migration of geothermal fluids (Sano and Wakita, 1985; Kennedy *et al*., 1987), (ii) evolution of hydrocarbon reservoirs (Ballentine *et al*., 1991; Hiyagon and Kennedy, 1992; Pinti and Marty, 1995; Kennedy *et al*., 2002), and (iii) pilot or analogue studies of geological $CO_2$ sequestration (Gilfillan *et al*., 2008, Gilfillan *et al*., 2009). Noble gas radionuclides add enhanced capability of determining the time scales of these processes and fluid migration. Furthermore, the rate of release of radiogenic and nucleogenic noble gases from the production sites in minerals to the fluid phase may also be determined uniquely through the studies of noble gas radionuclides (Yokochi *et al*., 2012), which is fundamental to the behavior of volatile elements in geochemistry.

In order to demonstrate proof of concept, a pilot study of noble gas radionuclides in an active geothermal system was performed at Yellowstone National Park (Yokochi *et al*., 2013). Prior studies of the Yellowstone system using stable noble gas isotopes show that the thermal fluids contain a mixture of atmospheric, mantle, and crustal components. The $^{39}Ar$ isotopic abundances in air-corrected samples exceeded those of atmospheric Ar by 525 to 1352%, indicating substantial contributions to the thermal fluids by the reservoir lithology. With the reasonable assumption that the fluid acquired its crustal component of Ar from the Quaternary volcanic rocks of the Yellowstone caldera, upper limits on deep thermal fluid mean residence times, estimated from $^{39}Ar/^{40}Ar^*$ ratios, ranged from 121 to 137 kyr for features in the Gibbon and Norris Geyser Basin areas, and are < 19.6 kyr in Lower Geyser Basin. $^{81}Kr$ isotopic abundances in the same Gibbon and Norris Basin samples are modern within analytical uncertainty, yielding upper limits on residence time that are consistent with those obtained from $^{39}Ar/^{40}Ar^*$ ratios.

### 5.4.3. Challenges in studying deep-seated fluids

Sampling deep-seated fluids is different from sampling groundwater, seawater and glacial ice in that the bulk sample delivered in the field is often in gaseous state and may be enriched in such species as $CO_2$ and $CH_4$ by more than two orders of magnitude. As a result, the total amount of gas to be sampled and processed has to increase by the same large factor. In an exploratory study, $Ca(OH)_2$ pellets were successfully used to remove $CO_2$ both in the field and in the laboratory (Yokochi *et al*., 2013). For $CH_4$ removal in Kr purification, the only established method is gentle combustion (see Section 4.2.1). Ongoing projects include studies of gases from $CO_2$-rich production and domestic wells.

Deep-seated fluids are prone to mixing with young meteoric components near surface during their ascent. In-situ produced isotopes from the reservoir rocks can also enter the fluids. Furthermore, multiple fluid components can be mixed in the subsurface. In order to more accurately interpret the noble-gas radionuclide results, the geochemical systems require a multi-tracer approach (e.g. Yokochi *et al*., 2013). The mixing of modern ambient air during sampling and young meteoric components may be quantified using $O_2$ concentration, the $^{85}$Kr/Kr ratio, and, ideally, another anthropogenic tracer (e.g. $^{3}$H or CFCs). After correcting the effects of in-situ production and mixing, the decrease in the abundances of noble-gas radionuclides along a flow path reflect the time of transport. Residence time of fluids in steady-state fluid reservoirs may also be determined if the noble-gas radionuclides are constantly supplied to the reservoir from groundwater flow or from the reservoir rock. The fluid "age" is model-dependent in all cases. Better understanding of the production rates and mechanisms of noble-gas radionuclides in reservoir rocks would significantly decrease the uncertainties in modeling fluid residence times.

# 6. Conclusions and Perspectives

It has been shown that in principle ATTA is well suited for measuring the rare noble gas radionuclides $^{85}$Kr, $^{39}$Ar and $^{81}$Kr. For $^{81}$Kr dating in the age range of 150 kyr – 1,500 kyr, the required sample size is 5 – 10 µL STP of krypton gas, which can be extracted from approximately 100 – 200 kg of water or 40 – 80 kg of ice. For $^{85}$Kr dating of young groundwater, the required sample size is generally a factor of 10 less. At present, Argonne National Laboratory and the University of Science and Technology of China each has one atom trap in operation and can analyze both $^{81}$Kr/Kr and $^{85}$Kr/Kr in ~ 120 samples per year.

Analysis of $^{39}$Ar/Ar in environmental samples has also been demonstrated. A group at Heidelberg University demonstrated an $^{39}$Ar atom count-rate of $4.1 \pm 0.3$ hr$^{-1}$ for atmospheric samples, which represents an improvement by a factor of 18 over what the Argonne group had demonstrated earlier. With only one day of data acquisition, the Heidelberg group dated a groundwater sample to be 360 ± 68 years. The instrument has the potential to measure $^{39}$Ar/Ar ratios in samples of 1 cm$^3$-STP of argon, which can be extracted from 100 cm$^3$ of air, 2.5 liters of water, or 1 kg of ice.

Gas extraction is performed routinely in the field using two types of systems: vacuum cylinder and membrane extractor, both having extraction efficiencies of 80-90%. Concentration and purification of argon and krypton from the bulk extracted gas is done in laboratory employing an array of techniques including cryogenic distillation, adsorption on molecular sieves and activated carbon, gas chromatography, and getter. These methods have proven to be robust although future application will certainly lead to refinements of instrumentation or extraction procedures.

Scientifically, ATTA measurements of $^{81}$Kr in the Nubian Aquifer of Africa, the Great Artesian Basin of Australia, and the Guarani Aquifer of South America have transformed our understanding of the long-term behavior of these large systems, allowing improved calibration and validation of numerical hydrodynamic models as well as cross-validation with other tracers ($^{4}$He and $^{36}$Cl). $^{85}$Kr is now routinely measured by ATTA and provides valuable complementary information to tritium-$^{3}$He, chlorofluorocarbons, and $SF_6$ as tracers for estimates of mean residence times in young (< 60 yr), typically shallow groundwater. We conservatively estimate the demand for $^{81,85}$Kr analysis by the U.S. hydrology community to be at least 1000 samples per year. With the recent success in ATTA analysis of $^{39}$Ar, this isotope will become a standard tool in the evaluation of groundwater resources, particularly because it addresses a critical time range (100-1,000 yr) presently not well resolved by other tracers

The few available $^{39}$Ar/Ar measurements performed by ultra-low level counting have clearly demonstrated the potential of $^{39}$Ar to provide better estimates of ocean ventilation, circulation and mean residence times of the principal water masses. Its half-life of 269 yr compares well to the mean residence times of the deep waters in the ocean and it does not have the reservoir effect of $^{14}$C. Combination of $^{39}$Ar and $^{14}$C will provide better constraints on circulation models. A systematic survey of $^{39}$Ar throughout the oceans would fill major gaps in our knowledge of deep ocean circulation and mixing, and would allow better predictions of the fate of atmospheric $CO_2$ if sequestered in the deep ocean. Sampling of roughly 1000 samples in the Pacific Ocean, 500 in the Indian Ocean, 200 in the Arctic Ocean and 300 repeat samples in the Atlantic Ocean would provide a reasonable one-time coverage. This adds up to a demand for measurement of ~2000 new samples and 1000 archived samples.

$^{81}$Kr could potentially be used for dating of old ice with ages ranging from 100 kyr – 1,500 kyr. Old ice can be obtained not only from deep ice cores, but also at ice margins and in Antarctic blue ice areas where it is being re-exposed by ablation. For paleoclimate studies this provides an interesting alternative to ice coring. The feasibility and accuracy of $^{81}$Kr-dating of old ice has been tested in an experiment using samples extracted from the well-dated stratigraphy of Taylor Glacier.

The fluid migration rate or storage time in crustal reservoirs has rarely been determined due to the lack of optimal tracers. The ability to routinely analyze crustal fluid samples for noble gas radionuclides ($^{39}$Ar, $^{81}$Kr and $^{85}$Kr) will provide an indispensable tool for deriving a thorough understanding of the origin, evolution and migration of crustal fluids, such as in geothermal, $CO_2$, and hydrocarbon reservoirs.

In order to utilize the full potential of ATTA, the following steps should be considered:

1. Establishment of a dedicated ATTA facility, with separate instrumentation for routine measurements and for research dedicated to the instrument and method development. The scope of such a facility would be dictated by the projected evolution of the ATTA technology, which currently allows 120 measurements of $^{81}$Kr and $^{85}$Kr per atom trap per year. An ATTA facility capable of analyzing all three noble-gas radionuclides ($^{85}$Kr, $^{81}$Kr, $^{39}$Ar) could be designed with multiple atom traps to optimize its efficiency and versatility.

2. Studies to optimize sample collection and purification methods should be continued. A standard design for a robust, portable gas extraction system would be desirable, along with a standard design for a separation line for purification of argon and krypton from extracted gas samples. Sample collection and purification protocols should be optimized for the sample-size requirements of the present generation of ATTA instruments.

3. Sampling strategies for noble gas radioisotopes on the basis of known properties of the study domain should be optimized. Due to the large effort involved in argon and krypton isotope measurements, some systems have been under sampled leading to ambiguities in the interpretation of the data. Increased capacity should help accommodate the increased sample load expected from optimized sampling plans.

## Acknowledgement

This paper is the outcome of a workshop supported by the Laboratory-Directed Research and Development Program of Argonne National Laboratory; the U. S. Department of Energy, Office of Nuclear Physics, under contract DEAC02-06CH11357; the U.S. National Science Foundation under grant EAR-1231372. We are grateful to all participants of the workshop for stimulating discussions, and to those who have made presentations and contributed materials based on which this article is written. The presenters include, in addition to the authors, W. Aeschbach-Hertig, P. Aggarwal, Ch. Buizert, M. Holzer, S.-M. Hu, W. Jiang, R. Kipfer, M. Kohler, W. Kutschera, A. Loose, A.J. Love, P. Mueller, T.M. Parris, F. Ritterbusch, C. Sukenik, and G. Winckler. A complete list of attendees is posted online at http://www.phy.anl.gov/events/tangr2012/ .